\documentclass{article}
\usepackage[pdftex]{graphicx}
\usepackage{natbib}
\usepackage[utf8]{inputenc}
\usepackage{authblk}

\usepackage{multirow}
\usepackage{hyperref}
\usepackage{indentfirst}
\usepackage{amsfonts, amssymb, amsmath, amsthm, bm} 
\usepackage{dsfont, accents}

\usepackage{float, enumerate}
\usepackage[mathscr]{euscript}

\usepackage{caption}
\usepackage{subcaption}
\usepackage[section]{placeins}

\title{Algorithmic and High-Frequency Trading Problems for Semi-Markov and Hawkes Jump-Diffusion Models}
\author[1]{Luca Lalor}
\author[2]{Anatoliy Swishchuk}
\affil[1]{\small University Of Calgary, Department of Mathematics and Statistics, University of Calgary, Calgary, AB T2N 1N4, Canada}
\affil[2]{\small University Of Calgary, Department of Mathematics and Statistics, University of Calgary, Calgary, AB T2N 1N4, Canada}

\date{\today}

\begin{document}
\maketitle
\thispagestyle{empty}

\begin{abstract}
This paper introduces a jump-diffusion pricing model specifically designed for algorithmic trading and high-frequency trading (HFT). The model incorporates independent jump and diffusion processes, providing a more precise representation of the limit order book (LOB) dynamics within a scaling-limit framework. Given that algorithmic and HFT strategies now dominate major financial markets, accurately modeling LOB dynamics is crucial for developing effective trading algorithms. Recent research has shown that LOB data often exhibit non-Markovian properties, reinforcing the need for models that better capture its evolution. In this paper, we address acquisition and liquidation problems under more general compound semi-Markov and Hawkes jump-diffusion models. We first develop jump-diffusion frameworks to capture these dynamics and then apply diffusion approximations to the jump components so that robust solutions can be given. Optimal trading strategies are formulated using stochastic optimal control (SOC) and solved numerically. Finally, we present strategy simulations analyzing price paths, inventory evolution, trading speed, and average execution prices. This study provides insights into how these models can improve execution strategies under more general price dynamics.
\end{abstract}


\pagenumbering{arabic}

\newpage

\section{Introduction}

The focus of this paper will involve formulating asset price processes of the jump-diffusion type, through a diffusion approximation for the jump parts. In general, we aim to develop stochastic midprice processes, which are referred to as the “fundamental price" in \cite{cartea2015algorithmic}, that portray an accurate representation of how the value of a financial asset evolves. In particular, we focus on assets where the majority of transactions take place within the LOB, thus the evolution of the price here would generally represent the value of these assets. As new information about these financial assets arrives into the market or certain transactions take place (the larger the more effect they tend to have), the fundamental price then evolves. This is often modeled as a combination of noise via the increments of Brownian motion along with some added drift that may be occurring, often based on certain traders actions. If the new information entering the market is very sudden and ground breaking, or the transactions taking place are abnormally large, big jumps can instantly occur in the fundamental price. In this case, a jump model would be necessary to accurately model the midprice dynamics. 

We will begin by defining a general version of the stochastic midprice processes used in \cite{cartea2015algorithmic}, as they regularly use this midprice process to solve Stochastic Optimal Control (SOC) trading problems. This midprice process can be defined as follows, 
\begin{align}
S_t = \pm g(\nu_t)t +\sigma W_t, \ S_0=S.
\label{eq: mp_cartea}
\end{align}
Here, $g:\mathbb{R}_+ \rightarrow \mathbb{R}_+$ is a function representing the permanent impact that a traders actions can have on the midprice of a financial asset. The $\pm$ before $g$ indicates whether the trading agent is buying ($+$) or selling ($-$) units of a financial asset. Assume, for now, $\nu_t$ represents some action a trader can take, like the buying or selling of a financial asset.  Lastly, $W_t$ is a standard brownian motion representing the diffusion dynamics and $\sigma$ is the diffusion coefficient.  


The models in this paper aim to extend and improve the stochastic midprice processes used for modeling LOB dynamics, under the SOC framework. High frequency LOB data often exhibit points of discontinuity i.e., jumps to different price levels, which a pure diffusion model would ignore. Jumps in the price process can occur for various reasons, such as unexpected news events that erode liquidity at nearby price levels or the sudden arrival of a large buyer or seller with high urgency to execute a trade. Thus, it is essential for a price process to incorporate these dynamics to more precisely characterize financial asset pricing behavior.

\begin{figure}[H]
	\centering
	\includegraphics[width=\linewidth]{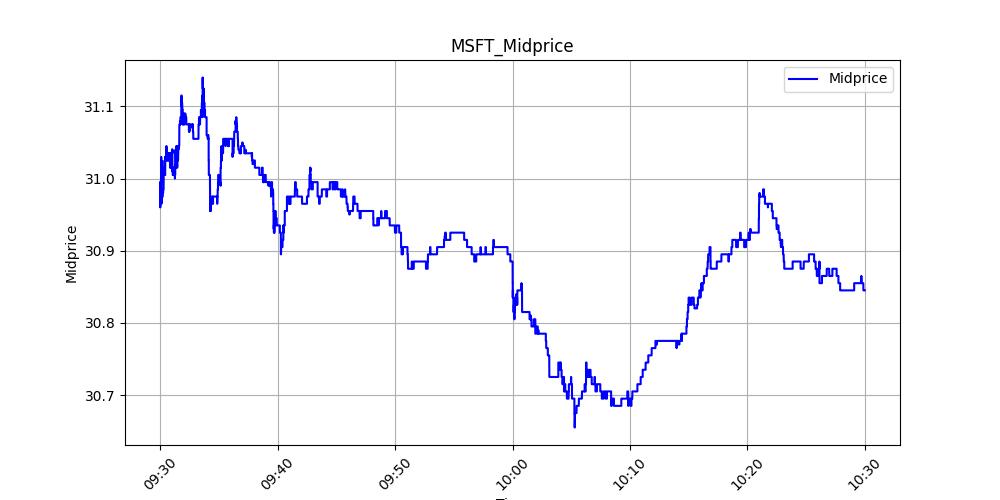}
	\caption{Midprice evolution in Microsoft's stock on June 21, 2012, from 9:30 AM to 10:30 AM Eastern Time (ET).}
	\label{fig:MSFT_midprice}
\end{figure}

Recent studies such as \cite{cont2012order}, \cite{swishchuk2017semi}, and \cite{kreher2023jump} have shown that jump-diffusion pricing dynamics is a good approximation technique for LOB data in financial markets. Upon visualizing the data, as can be done by looking at Figure \ref{fig:MSFT_midprice}, one would notice that it appears to closely resemble diffusion dynamics, but at some points the data also experience sharp jumps. In addition to exhibiting jump dynamics, LOB data, in general, tend to follow non-Markovian dynamics as shown by many extensive studies on the data. The empirical results in \cite{he2019quantitative}, \cite{swishchuk2019compound} and \cite{swishchuk2020general} show that higher numbers of state-dependent orders are often more accurate in modeling LOB midprice dynamics versus the mathematical assumption of an infinitesimal tick size in a general GBM model. Similar evidence can be found in many more works and some examples, in no particular order, include \cite{cartea2018enhancing}, \cite{cartea2018algorithmic}, \cite{drame2019limit}, \cite{makinen2019forecasting}, \cite{sfendourakis2021lob}, \cite{sjogren2021general},  and many more. Upon examining the LOB data, these papers all build models with the underlying assumption that jumps in price occur. 

Thus, our aim in this paper is to add more versatility to the pricing models used to formulate and solve algorithmic trading problems, as seen in \cite{cartea2015algorithmic}, where they mostly use midprice processes based on a pure arithmetic diffusion model, as shown in Equation \eqref{eq: mp_cartea}. Therefore, our models are new and more general for both the cases considered in: i) \cite{cartea2015algorithmic} for diffusion case only with drift, see (1),  and ii) in \cite{roldan2022optimal} and \cite{roldan2023stochastic}, where the pricing models only considered jumps by applying diffusion approximations for the Semi-Markov LOB models in \cite{swishchuk2017semi} and the General Compound Hawkes Process (GCHP) LOB models in \cite{swishchuk2020general}. We aim to extend both cases, the first one with only diffusion and the second one with only jumps, by formulating more general price processes to account for these dynamics. 

The main contributions of this paper can be summarized as follows.
\begin{enumerate}
    \item We propose a jump-diffusion pricing model tailored for algorithmic and HFT, integrating independent jump and diffusion processes that more accurately capture LOB dynamics under a scaling-limit approximation. Our framework uniquely allows the jump dynamics to be modeled using either a Semi-Markov or Hawkes process, combined with an independent Brownian motion component. While a pure jump model based on Semi-Markov and Hawkes dynamics in algorithmic and HFT settings is not new, the incorporation of standardized non-Markovian jump components alongside a conventional arithmetic Brownian motion via Jump-Diffusion models in practical applications represents a significant advancement.
    \item To facilitate more robust solutions for standard algorithmic and high-frequency trading (HFT) models in limit order book modeling, where transactions occur every millisecond, we provide approximations for the jump components in our jump-diffusion pricing models. In applications, one typically considers a longer time scale, $tn$ (minutes, hours, etc.), rather than $t$ (milliseconds), allowing sufficient time for solving standard algorithmic and HFT trading problems. In this context, limit theorems such as the law of large numbers (averaging) and the functional central limit theorem (diffusion approximation), which apply in the limit as $n \to \infty$ while $t$ remains fixed, become particularly useful.
    \item Using real LOB data, we used calibrated model parameters and conducted an in-depth numerical analysis of both acquisition and liquidation problems under our proposed pricing dynamics. This was followed by extensive simulations across various scenarios where the jump component varies significantly, capturing different market regimes. Our analysis demonstrates how the solutions and simulation outcomes can vary considerably in response to these shifting jump dynamics. 
\end{enumerate}

The paper is organized as follows. Section 2 introduces jump-diffusion models in HFT, namely, semi-Markov and Hawkes process ones. Section 3 studies two stochastic optimal control trading problems in algorithmic and HFT, namely an acquisition and liquidation problem, under our new jump-diffusion models. Numerical solutions to these acquisition and liquidation problems are considered in Section 4. Strategy simulations for these acquisition and liquidation problems are presented in Section 5. Lastly, our conclusions and future recommendations are discussed in Section 6. 
\sloppy
\section{Semi-Markov and Hawkes Jump-Diffusion Models in HFT}
\fussy

This Section is devoted to the modeling of Jump-Diffusion pricing models, whereby the jump parts are modeled via Semi-Markov and Hawkes process dynamics. In \cite{cartea2015algorithmic}, Markovian jump-diffusion pricing models are mentioned with a diffusion part $W(t)$ and a counting process $N(t)$. However, these models lack robustness when applied to algorithmic and HFT problems, making them unsuitable for practical implementation. Moreover, no existing approach has been proposed for addressing more general jump-diffusion frameworks, such as those involving Compound Semi-Markov and Compound Hawkes processes, in real-world applications. Subsequently, here we propose to approximate the jump parts of these price processes to obtain a diffusion approximation, allowing our more general compound Semi-Markov or compound Hawkes processes to be included in trading problem solutions. Thus, this work recognizes that trading agents often face multiple risk factors, as also identified in \cite{cartea2015algorithmic}, and proposes a more comprehensive modeling framework to effectively account for these factors in practical applications.

An important consideration with this type of pricing model, as seen in Equation \eqref{eq: mp_cartea} from \cite{cartea2015algorithmic} and later in our own models, is its potential drawback: the possibility of negative prices, a limitation that can be further exacerbated by the presence of jumps. To address this, the pricing model should ensure that initial prices are much higher than the potential jump tick sizes, where an extreme amount of jump movements in one direction would have to occur for prices to go negative. As we will show in our simulations later, the price process remains well above negative values, even under extreme scenarios. To fully mitigate this drawback and eliminate any ambiguity, this issue could be addressed by adopting an exponential jump-diffusion pricing model. A similar approach was formulated for the pure jump case in \cite{guo2022multivariate} within a limit order book (LOB) framework. 

The remainder of this Section will proceed as follows. In Section 2.1, we introduce our jump-diffusion model under our Semi-Markov approximation approach (via Law of Large Numbers) for the jump part. In Section 2.2, we similarly introduce our jump-diffusion model but under the Hawkes process approximation approach (via a functional central limit theorem) for the jump part. 

\subsection{Jumps as a function of a Semi-Markov Process}

To first give some background, previous models, such as in \cite{cont2013price}, suggest that for modeling the LOB, one can use a straightforward stochastic model for the dynamics of an LOB, in which the arrivals of market orders, limit orders, and cancellation orders are described in terms of a Markovian queuing system. In \cite{swishchuk2017semi}, this work was extended to reflect the fact that an arbitrary distribution for the inter-arrival times of orders inside the LOB tend to be non-exponential. They portray how the nature of a new book event and its corresponding inter-arrival times depend on the nature of the previous book event. This contradicts the Markovian model in \cite{cont2013price} and the use of a Poisson and Compound Poisson process to model the jump part of the stock price process. To improve the model to reflect this non-Markovian property, \cite{fodra2015semi} uses a Markov Renewal process, also known as a Semi-Markov process, with this paper focusing on this from a market microstructure perspective. In \cite{fodra2015high}, they study an optimal high frequency trading problem where the price process is driven by a Markov Renewal Process. 

Semi-Markov processes are formed from a wide array of stochastic processes. The main advantage over a regular Markov process is that it allows the use of an arbitrary waiting time distribution for modeling the time until a transition occurs out of one state to another i.e., waiting times do not have to be modeled using an exponential distribution. Past work first developed a General Semi-Markov model (in \cite{swishchuk2017general}) for modeling the dynamics of LOBs and then \cite{swishchuk2017semi} developed an approximation to the Semi-Markov process as a diffusion model. In \cite{roldan2023stochastic}, this led to an additional drift term in the stock price process when they solved SOC algorithmic trading problems. In this paper, we first formulate a model for the case where the price process follows the regular diffusion model from Equation \eqref{eq: mp_cartea} plus a jump part, i.e., a jump-diffusion model, as follows:
\begin{align}
S_t = S_0 \pm g(\nu_t)t + \sigma W_t + \sum_{k=1}^{N(t)} X_k
\label{eq:mp_SM_jumpdif}
\end{align}
In Equation \eqref{eq:mp_SM_jumpdif}, $W_t$, $\sigma$ and $g$ are as defined before, $X_k$ is an ergodic Markov chain representing two types of consecutive price increments taking values $\pm \delta$ and $N(t)$ represents the number of renewals (or arrivals) that have occurred up to time $t$ i.e., $N(t)$ is the counting process for the price changes. This price process now includes an additional jump term, $\sum_{k=1}^{N(t)} X_k$, which is not included in Equation \eqref{eq: mp_cartea}.  $N(t)$ is a counting process independent of $X_k$ and $W_t$ i.e., all sources of randomness are assumed to be independent. This means that the number of price changes, $N(t)$, does not depend on the tick size described by $X_k$. Thus, we can gain more insight into how an SOC algorithmic trading problem would perform using a price process that includes the dynamics of this jump part.  

Next, we will introduce the first of our proposed diffusion approximation models for the jump part of the price process in Equation \eqref{eq:mp_SM_jumpdif}. This, along with the rest of the regular diffusion model, will enable us to solve some of the algorithmic and HFTs problem from \cite{cartea2015algorithmic} using this more innovative price process that accounts for the non-markovian jump dynamics. For this purpose, we will utilize the diffusion approximation in \cite{swishchuk2017semi} under the assumption (A5), sec. 4.2., of a finite tick size. Then the price process in Equation \eqref{eq:mp_SM_jumpdif} becomes,
\begin{align}
\begin{split}
S_t &= S_0 + (\pm g(\nu_t)\eta_{SM})t + \sqrt{\sigma^2 + \bar{\sigma} ^2_{SM} + \left(\frac{\sigma^*}{\sqrt{m_{\tau}}}\right)^2}W_t \\&= S_0 + (\pm g(\nu_t)\eta_{SM})t + \sqrt{(\sigma^2 + \bar{\sigma}^2_{SM}+ \varsigma_{SM}^2)}W_t
\label{eq:mp_SM_difapprox}
\end{split}
\end{align}
where here the drift coefficient, $\eta_{SM}$, can be represented, as in \cite{roldan2023stochastic}, for the balanced market case as,
\begin{align}
\eta_{SM} = \frac{1}{m_{\tau}}s^*
\label{eq:eta_SM}
\end{align}
where the SM index refers to the fact that this coefficient originates from a Semi-Markov process. For the Hawkes process case, we will introduce a similar coefficient, $\eta_{HP}$, thus, we index this coefficient to avoid confusion later on. Here, $m_{\tau} = \sum_{i\in\{-\delta,\delta\}} \pi^*(i)m(i)$ and the normalized prices $s^*$ is defined as $s^*:=\delta(2\pi^*-1)$, where $\pi^*$ is a long-run probability. In the diffusion approximation, we use the scaling limit,  $\lim_{n \to \infty}$, for $S_{tn}$. 
For a complete proof to show that the diffusion part of this renormalized price process satisfies a weak convergence in the Skorokhod topology, see \cite{swishchuk2017semi}. In applications, such as algorithmic and HFT, and in LOB modeling
(when transactions happen every milliseconds), one usually considers the long
scale $tn$ (minutes, hours, etc.) instead of $t$ (milliseconds) to have time to prepare
solutions for trading problems such as liquidation, acquisitions, and market making,
to name a few, where in this paper we will study the first two. In this case, the limit theorems that use the law of large numbers
(averaging) for the case where $n$ goes to infinity and $t$ is fixed are very useful.

Following along in a similar fashion, $\varsigma_{SM} = \frac{\sigma^*}{\sqrt{\tau}}$ is another new coefficient representing the Semi-Markov diffusion approximation for the jump part of Equation \eqref{eq:mp_SM_jumpdif}. Here, $\sigma^*$ is a constant depending on the ergodic and transition probabilities of the Markov chain $X_k$ and $\tau$ refers to the interarrival time of the jumps. Then for our two state Markov chain, $\sigma^*$ can be defined, as in \cite{swishchuk2017semi}, as,
\begin{align}
(\sigma^*)^2 = \sqrt{4\delta^2\left(\frac{1-p_{cont}^{\text{'}}+\pi^*(p_{cont}^{\text{'}}-p_{cont})}{(p_{cont}+p_{cont}^{\text{'}}-2)^2}\right)},
\label{eq:sig_star_SM}
\end{align}
where $p_{cont} = P[X_{k+1} = \delta | X_k = \delta]$, $p_{cont}^{\text{'}} = P[X_{k+1} = -\delta | X_k = -\delta]$, $\pi^* = \frac{p_{cont}^{\text{'}}-1}{p_{cont}+p_{cont}^{\text{'}}-2}$, $\tau = \sum_{k=1}^{\infty} \sum_{p=1}^{\infty} \alpha^b(k)\alpha^a(p)f^*(k,p)$ and $f^*(k,p) = \pi^*f(k,p)+(1-\pi^*)\tilde{f}(k,p)$, where $f(k,p)$ is the probability distribution after a price increase and $\tilde{f}(k,p)$ is the probability distribution after a price decrease and $\alpha$ refers to the function of the interarrival times, where the exponents $a$ and $b$ refer to the ask and bid sides of the LOB, respectively (see  \cite{swishchuk2017semi} for more details). Lastly, $\bar{\sigma}_{SM}$ can be as defined in \cite{swishchuk2017semi} as follows:
\begin{align}
\bar{\sigma}_{SM} = \sqrt{\frac{(\sigma^*)^2}{m_{\tau}}+\frac{\Pi\sigma^2}{m_{\tau}}},
\label{eq:bar_sig_SM}
\end{align}

To extend these results to a larger number of states, see the results in \cite{swishchuk2017semi}, where one can define this for a $n$-state dependent Markov chain. One can also find more details here on the Semi-Markovian modelling of LOBs. Here detailed proofs are given to show that this method for normalizing the price process holds following weak convergence in the Skorokhod topology for a $n$-state Markov chain. 

\subsection{Jumps as a function of a Hawkes Process}
To first give a little background, the Hawkes process was first introduced by its creator Alan Hawkes in \cite{hawkes1971spectra} as a self-exciting process and he expanded on this theory in \cite{hawkes1974cluster} by discussing how this can be applied to problems in which the data exhibits a clustering effect. A one dimensional Hawkes process is a point process $N(t)$ which is characterized by its intensity $\lambda(t)$ with respect to its natural filtration,
\begin{align}
\lambda (t) = \lambda + \int_0^t \mu(t-s)dN(s),
\label{eq:int_func_HP}
\end{align}  
where $\lambda>0$ is the background intensity and the response function $\mu(t)$ is a positive function, often referred to as the excitation function, which satisfies $\int_0^{\infty}\mu(s)ds<1$. In more recent times, Hawkes processes have been applied to problems in finance, as explained in \cite{bacry2015hawkes}, where certain processes incur these self-exciting and clustering effects. These properties are also more consistent when modelling price processes with jumps in LOBs, which the empirical results show in \cite{fodra2015semi}. We, of course, in this paper specifically focus on their applications to modeling LOBs. Some examples of previous work related to modeling LOBs include \cite{swishchuk2020general}, where they introduced the General Compound Hawkes Process (GCHP) in LOBs with an application in a SOC setting given in \cite{roldan2022optimal}, or in \cite{cartea2018algorithmic}, where they use HPs to try and predict adverse selection within the LOB, also with applications under the SOC framework.

Here, like in Section 2.1., we model the price process with a diffusion and jump part as follows:
\begin{align}
S_t = S_0 \pm g(\nu_t)t + \sigma W_t+\sum_{k=1}^{N(t)} a(X_k),
\label{eq:mp_HP_jumpdif}
\end{align}
where $g$ and the diffusion term are as defined before in Section 2.1. In Equation \eqref{eq:mp_HP_jumpdif}, $a$ maps the current state to a price movement, $X_k$ is again an ergodic Markov chain with ergodic probabilities $(\pi_1^*, \pi_2^*,...,\pi_n^*)$, where the current state is computed using information from the previous state. Here, $N(t)$ is again a counting process independent of $X_k$ and $W_t$. Then using the diffusion approximation for the jump part in \cite{swishchuk2020general}, this becomes,
\begin{align}
\begin{split}
S_t &= S_0 \pm (g(\nu_t)\eta_{HP})t+ \sqrt{\sigma^2 + \bar{\sigma} ^2_{HP} + \left(\sigma^*\sqrt{\frac{\lambda}{1-\hat{\mu}}}\right)^2}W_t,\ S_0=S \\& = S_0  \pm (g(\nu_t)\eta_{HP})t+ \sqrt{\sigma^2+\bar{\sigma}^2 _{HP} + \varsigma_{HP}^2}W_t,\ S_0=S.
\end{split}
\label{eq:mp_HP_difapprox}
\end{align}
The scaling limit for the diffusion approximation part of the midprice process in Equation \eqref{eq:mp_HP_difapprox} is formulated via the functional central limit theorem (FCLT) for the General Compound Hawkes process in \cite{swishchuk2020general}, where as before in Section 2.1, we use the scaling limit, $\lim_{n \to \infty}$, for $S_{tn}$, which holds 
in a weak sense in the Skorokhod topology. A proof of this claim can be found in \cite{swishchuk2020general}. Similarly to the argument given in Section 2.1, the usefulness in Equation \eqref{eq:mp_HP_difapprox} is in terms of being able to prepare solutions to algorithmic and HFT trading problems, where the FCLT enables the use of a diffusion approximation, for the case when $n$ goes to infinity and $t$ is fixed.
Here, the drift coefficient can be defined, as in \cite{roldan2023stochastic}, as,
\begin{align}
\eta_{HP} = a^*\frac{\lambda}{1-\hat{\mu}},
\label{eq:eta_HP}
\end{align}
where $a^*$ is a constant depending on the transition probabilities of the Markov chain. This can be defined, as in \cite{swishchuk2020general}, as $a^*:= \sum_{i \in X}\pi_i^*a(i)$, where $\pi^*$ are the ergodic probabilities of the Markov chain $X$. $\lambda$ is again the background intensity and $\mu$ refers to the output of the response/excitation function. Here, $\varsigma_{HP} = \sigma^*\sqrt{\frac{\lambda}{1-\hat{\mu}}}$. $\sigma^*$ is again a constant depending on the ergodic probabilities of the Markov chain $X_k$. Since this kind of Markov chain has two states, where $X_k \in \{-\delta,\delta\}$, $\sigma^*$ can be defined, as shown by corollary 3 in \cite{swishchuk2020general}, as
\begin{align}
\sigma^*:= \sqrt{4\delta^2\left(\frac{1-p^{\text{'}}+\pi^*(p^{\text{'}}-p)}{(p+p^{\text{'}}-2)^2}-\pi^*(1-\pi^*)\right)}.
\label{eq:sig_star_HP}
\end{align}
Also, note that $\bar{\sigma}_{HP}$ can be defined as in \cite{swishchuk2020general} as,
\begin{align}
\bar{\sigma}_{HP} = \sqrt{(\sigma^*)^2+\left(a^*\sqrt{\frac{\lambda}{1-\hat{\mu}}}\right)},
\label{eq:barsig_HP}
\end{align}
where,
\begin{align*}
\sigma^* = \hat{\sigma}\sqrt{\frac{\lambda}{1-\hat{\mu}}}\ \text{and} \ \hat{\sigma}^2 := \sum_{i \in X}\pi_i^*v(i), 
\end{align*}
where $v(i)$ represent the transitions in and out of states. Lastly, in Equation \eqref{eq:mp_HP_difapprox}, $\lambda$ and $\hat{\mu}$ represent the background intensity and the excitation/response function, respectively. Recall, more formally, $\hat{\mu}$ can be defined, as also stated in \cite{swishchuk2020general}, as,
\begin{align}
0<\hat{\mu}:=\int_0^{\infty}\mu(s)ds<1\ and \int_0^{\infty} s\mu(s)ds<\infty.
\end{align}

\section{Stochastic Optimal Control Trading Problems: Acquisition and Liquidation}

SOC theory is utilized in a variety of different continuous-time problems to optimize the performance of a particular model. Some of the first historical examples in finance were developed in the 1970s in the seminal paper by \cite{merton1969lifetime} which is devoted to the portfolio allocation problem. The results of this model were then extended by many others, such as \cite{zariphopoulou1989optimal} to optimal investment-consumption models, \cite{davis1990portfolio} to portfolio selection with transaction costs, and \cite{oksendal2002optimal} to the portfolio selection problem with fixed and variable transaction costs. In this paper, we will focus on applying the SOC framework to algorithmic and HFT problems, where a whole range of examples can be found in \cite{cartea2015algorithmic}, one of which we will solve with our new price processes from Section 2. 

In general, the agent in our examples would either like to acquire/buy or liquidate/sell a large number of units (shares, lots, etc) over a certain time horizon, say from $[0, T]$, where it would be sub-optimal to acquire/liquidate the targeted number of units instantly.  These types of trading scenarios occur regularly and particularly when the total amount to be acquired or liquidated is much larger than the available liquidity in the LOB. The general setup of these types of trading problems involves using stochastic optimal control theory, where the goal is to maximize/minimize an objective function by acting in an optimal manner. Thus, the agent can regulate the specific set of actions it can take, which is referred to as the control. The control variable in the case of either the acquisition or liquidation problem we study is the speed/rate of trading. The optimal control is then driven by the dynamics of the asset price and many other processes (deterministic or stochastic) deemed relevant to the problem. The key stochastic processes in these types of trading problems, as described in \cite{cartea2015algorithmic}, are as follows, 
\begin{itemize}
\item $\nu=(\nu_t)_{\{0 \le t \le T\}}$ is the trading speed/rate, the speed at which the agent is buying or selling units of an asset. This is the variable that the agent can control.
\item $Q^{\nu}=(Q_t^{\nu})_{\{0 \le t \le T\}}$ is the agents' inventory. This is impacted by the speed at which the agent trades, hence this is the agent's controlled inventory process. 
\item $S^{\nu}=(S_t^{\nu})_{\{0 \le t \le T\}}$ is the midprice process. This is also affected by the speed of the agent’s trades and so is the agent's controlled price process.
\item $\hat{S}^{\nu}=(\hat{S}_t^{\nu})_{\{0 \le t \le T\}}$ is linked to the price process at which the agent can acquire/liquidate the asset i.e., the execution price, by walking the buy/sell side of the LOB.
\item $C^{\nu}=(C_t^{\nu})_{\{0 \le t \le T\}}$ is the agent's cash process resulting from the execution strategy. 
\end{itemize}

Next, we see how these processes satisfy certain differential equations, which may be stochastic, as follows:
\begin{itemize}

\item The agent’s controlled inventory process is given in terms of her trading rate as,\begin{align} 
dQ_t^{\nu}=\pm v_tdt, Q_0^{\nu}=q,
\label{eq:inv_process}
\end{align}
where we use $+$ within an acquisition problem as the agents trading speed then has a positive relationship with inventory and $-$ in a liquidation problem as the agents trading speed then has a negative relationship with inventory. 
\item Following the analysis in Section 2, the midprice process can be defined for the Semi-Markov case as follows,
\begin{align}
dS_t^{\nu} = \pm(g(\nu_t)\eta_{SM}) dt + \sqrt{\sigma^2+\bar{\sigma}^2_{SM} + \varsigma_{SM}^2} dW_t,\ S_0^{\nu}=S,
\label{eq:dif_eq_mp_SM}
\end{align}
and for the Hawkes process case as,
\begin{align}
dS_t^{\nu} = \pm(g(\nu_t)\eta_{HP}) dt + \sqrt{\sigma^2+\bar{\sigma}^2_{HP} + \varsigma_{HP}^2} dW_t,\ S_0^{\nu}=S. 
\label{eq:dif_eq_mp_HP}
\end{align}
Here, $g, \eta_{SM}, \eta_{HP}, \varsigma_{SM}, \varsigma_{HP}, \sigma$, $\bar{\sigma}^2_{SM}$, $\bar{\sigma}^2_{HP}$ and $W$ are as defined in Section 2. The permanent price impact function, $g$, is positive in an acquisition problem as the actions of the agent create an upward drift in the price and negative in a liquidation problem as the actions of the agents create a downward drift in the price. Some of the foundational work on price impact models includes \cite{bertsimas1998optimal}, which focuses on optimizing trade execution costs given a price impact function. In addition, \cite{almgren2001optimal} and \cite{colaneri2020optimal} introduce permanent and temporary market impact as forms of transaction costs in their trading models. The latter also addresses an optimal liquidation problem, which is one of the trading problems examined in this study. Next, we will rewrite this midprice process one more time so that we have just one price process for both cases, and we will proceed with this formulation when solving the acquisition and liquidation problems later on. This price process will be defined as,
\begin{align}
dS_t^{\nu} = \pm(g(\nu_t)\eta) dt + \sqrt{\sigma^2+\bar{\sigma}^2 + \varsigma^2} dW_t, 
\label{eq:mp_nosubscript}
\end{align}
where we have now dropped the index referencing the Semi-Markov and Hawkes process parts of the coefficients $\eta$, $\bar{\sigma}$ and $\varsigma$, as the rest of the mathematical formulations in this paper are identical for both cases. 

\item The execution price process satisfies,
\begin{align}
\hat{S}_t^{\nu}=S_t^{\nu}+\displaystyle\left(\frac{1}{2} \Delta \pm f(\nu_t)\right), \hat{S}_0^{\nu} =\hat{S}.
\label{eq:exec_price}
\end{align}
Here, as in \cite{cartea2015algorithmic}, $f: \mathbb{R}_+ \rightarrow \mathbb{R}_+$ represents the temporary price impact that the agent’s actions have on the price at which they execute their trades at and $\Delta \ge 0$ is the bid-ask spread, which will here assumed to be constant. This temporary impact, similar to the permanent impact, is positive in an acquisition problem and negative in a liquidation problem. 

\item The agents cash process satisfies the differential equation,
\begin{align}
dC_t^{\nu}=\hat{S}_t^{\nu}\nu_tdt, C_0=c.
\label{eq:cash_process}
\end{align}

\end{itemize}

Next, we will first apply our pricing models to the acquisition trading problem with a price limiter from \cite{cartea2015algorithmic} in Section 3.1. Here, there will be a price cap on the price the agent can acquire units. Then, in Section 3.2, we proceed to develop a similar type of trading problem, but in the form of a liquidation problem, with a price floor on the price the agent can liquidate units. Recall that a trading agent or any rational investor aims to buy low and sell high. Thus, the price cap in the acquisition problem is there to avoid acquiring units at prices that are too high and similarly, the price floor in the liquidation problems, aims to avoid liquidating units at prices that are too low. Our general aim here is to show how our price processes can be applied within these types of trading problems. 

\subsection{Acquisition Problem}

In this trading problem, as explained in \cite{cartea2015algorithmic}, the agent’s objective is to acquire $\mathfrak{N}$ units over a trading horizon $T$, with a cap on the price at which the agent acquires units equal to $S_{max}$. The agent will stop trading if the agent has acquired $\mathfrak{N}$ units, the terminal time $T$ is reached, or the midprice $S_t$ (given by Equation \eqref{eq:mp_nosubscript}) with a positive drift term) has reached the price cap $S_{max}$. To formalize this, define a stopping time as
\begin{align}
\tau = T \wedge \{t:S_t^{\nu}=S_{max}\} \wedge \{t:Q_t^{\nu}=\mathfrak{N}\}.
\label{eq:stop_time_Acq}
\end{align}

When the terminal time $T$ or the price cap $S_{max}$ is reached, the agent purchases the remaining $\mathfrak{N}-Q_\tau^v$ units and pays $S_\tau+\alpha(\mathfrak{N}-Q_\tau^v)$ per share, where $\alpha$ represents the terminal acquisition penalty. One can also define the remaining units to be purchased by $Y_t^{\nu}= \mathfrak{N}-Q_\tau^{\nu}$, satisfying,
\begin{align}
dY_t^{\nu}=\nu_tdt
\label{eq:remaining_inv_acq}
\end{align}
where $\nu_t$ is the positive rate of trading.

The agents’ performance criteria, as in \citep{cartea2015algorithmic}, is defined as follows,
\begin{align}
H^{\nu}(t,S,y) = \mathbb{E}_{t,s,y}\displaystyle{\left[\int_t^{\tau}(S_u+f(\nu_u))du+y_{\tau}(S_{\tau}+\alpha y_{\tau})+\phi \int_t^{\tau}y_u^2du\right]}
\label{eq:perf_crit_Acq}
\end{align}
Here, $\phi \int_t^{\tau}y_u^2du$ with $\phi \ge{0}$ is a running inventory penalty of the remaining units to be acquired. The value function is next defined as
\begin{align}
H(t,S,y) = \inf_{\nu \in \mathscr{A}}H^{\nu}(t,S,y), \forall 0 \le t \le T, S  \le S_{max}, 0 \le y \le Q
\label{eq:val_func_Acq}
\end{align}
Here, $\mathscr{A}$ is the set of admissible strategies in which $\nu >0$ and uniformly bounded from above. 

Set $g(\nu)=b\nu$ and $f(\nu)=\kappa\nu$ as in \cite{cartea2015algorithmic} and  assume that our permanent and temporary impact functions are linear in the speed of trading, where $b \ge 0$, and $k>0$ are finite constants. Then by applying the DPP, the value function should satisfy the following DPE,
\begin{align}
\begin{split}
&\partial_t H(t,S,y) + \frac{1}{2}(\sigma^2 +\bar{\sigma}^2 +  \varsigma^2)\partial_{ss} H(t, S, y)+\phi y^2\\& + \inf_{\nu \in \mathcal{A}}\{-\nu\partial_yH(t,S,y)+b\nu\eta\partial_yH(t,S,y)+(S+\kappa\nu)\nu\}=0,
\end{split}
\label{eq:DPE_Acq}
\end{align}
subject to the terminal and boundary conditions:
\begin{align}
H(T,S,y)=(S+\alpha y)y;H(t,S_{max},y)=(S_{max}+\alpha y)y;H(t,S,0)=0
\label{eq:TC_BC_Acq}
\end{align}
Here, remember the coefficients $\eta$, $\bar{\sigma}$ and $\varsigma$ are representing either the cases for $\eta_{SM}$, $\bar{\sigma}_{SM}$ and $\varsigma_{SM}$, or $\eta_{HP}$, $\bar{\sigma}_{HP}$ and $\varsigma_{HP}$, from Equations \eqref{eq:mp_SM_difapprox} and \eqref{eq:mp_HP_difapprox}, respectively. One can see from the terminal and boundary conditions, that once the stopping time (see Equation \eqref{eq:stop_time_Acq}) is reached, the agent will purchase the remainder of the units. Also, one can see that once the agent has purchased the $\mathfrak{N}$ units initially targeted, the agent stops trading and there is no penalty, hence, the boundary condition $H(t,S,0)=0$ where $y=0$. 

Then, one can obtain the optimal trading speed, by using the mathematical procedure called completing the squares on the inside of the minimization part of Equation \eqref{eq:DPE_Acq}. And so, the optimal trading speed in feedback form can be defined as,  
\begin{align}
\nu^*(t,S,y) = -\frac{1}{2\kappa}(b\eta\partial_sH(t,S,y)-\partial_yH(t,S,y)+S).
\label{eq:Opt_control_Acq}
\end{align}
One can then substitute this optimal control back into the DPE in Equation \eqref{eq:DPE_Acq} and we then get the PDE:
\begin{align}
\begin{split}
&\partial_tH(t,S,y)+\frac{1}{2}(\sigma^2+\bar{\sigma}^2 + \varsigma^2)\partial_{SS}H(t,S,y)\\ &-\frac{1}{4\kappa}(b\eta\partial_s H(t,S,y)-\partial_y H(t,S,y) + S)^2 + \phi y^2 = 0
\end{split}
\label{eq:PDE_Acq}
\end{align}
This is in the same format as in \citep{cartea2015algorithmic}, except now we include the extra drift and diffusion coefficients from \cite{roldan2023stochastic} and we have added an additional diffusion coefficient, $\varsigma$, representing a more general version of the diffusion approximation of the jump parts, for either the Semi-Markov or Hawkes process cases.

Next, as in \cite{cartea2015algorithmic} we reduce the dimensions in order to be able to solve this problem numerically. In order to do so, we first set $b=0$ using their same argument that the permanent impact of walking the LOB tends to be relatively small compared to the temporary impact. This point is also emphasized in \cite{almgren2001optimal}. Although this simplifies the modeling approach, one slight drawback is that the model could underestimate any long-term market impacts that may occur. And so, we can use the following ansatz,
\begin{align}
H(t,S,y) = yS + y^2 h(t,S)
\label{eq:ansatz_acq}
\end{align}
and we see that $h$ satisfies the following PDE,
\begin{align}
\partial_t h(t,S)+\frac{1}{2}(\sigma^2 + \bar{\sigma}^2 + \varsigma^2)\partial_{SS}h(t,S)-\frac{1}{\kappa}h^2 + \phi = 0.
\label{eq:pde_h_acq}
\end{align}
The existence of solutions to this type of PDE follow from the same case as in \cite{cartea2015algorithmic} in the general case, which can be proven to be guaranteed under the so-called verification theorem following along the lines of many popular works such as \cite{pham2009continuous} and \cite{yong2012stochastic}. This PDE now has a new terminal and boundary condition defined as,
\begin{align}
h(T,S) = \alpha, S\le S_{max},\label{eq:TC_h_Acq}\\
h(t,S_{max}) = \alpha, t \le T.
\label{eq:BC_h_Acq}
\end{align}
The optimal control for this acquisition problem, $\nu^*$, can now be defined as,
\begin{align}
\nu^*(t,S) = \frac{1}{\kappa}yh(t,S).
\label{eq:control_h_acq}
\end{align}
Intuitively, Equation \eqref{eq:control_h_acq} states that the optimal acquisition speed decreases as the inventory Q increases, which would make sense as you are then closer to your target inventory.  

Lastly, as in \cite{cartea2015algorithmic}, we would like to portray how the current status of the inventory process Q can be obtained in terms of the path of the midprice process by computing $Y_t$, the process which determines how many more units are left to purchase. Here, $Y_t$ satisfies the SDE,
\begin{align}
dY_t^* = -\nu^*dt = -\frac{1}{\kappa}Y_th(t,S_t)dt,
\label{eq:remaining_inv_h_acq}
\end{align}
and therefore, 
\begin{align}
Q_t^* = \left(1-e^{ -\frac{1}{\kappa}\int_0^th(u,S_u)du}\right)\mathfrak{N}, \ t \le \tau.
\label{eq:opt_inv_h_acq}
\end{align}
Here, we can see that the midprice path taken is paramount in computing the level of inventory the agent has acquired at any given time. In Section 5, we will show some examples, through strategy simulations, how the optimal strategy performs over different midprice paths.

\subsection{Liquidation Problem}

In this trading problem, the agent’s objective is to liquidate $\mathfrak{N}$ units again over a trading horizon $[0,T]$, where now there is a floor on the price at which the agent liquidates units equal to $S_{min}$. This setup is similar to the acquisition problem in Section 3.1 but viewed from a liquidation perspective. The agent will stop trading if the agent has liquidated $\mathfrak{N}$ units, the terminal time $T$ is reached, or the midprice $S_t$ has hit the price floor $S_{min}$. To formalise this, we define a stopping time as
\begin{align}
\tau = T \wedge \{t:S_t^{\nu}=S_{min}\} \wedge \{t:Q_t^{\nu}=0\}. 
\label{eq:stop_time_liq}
\end{align}
When the terminal time $T$ or the price floor $S_{min}$ is reached, the agent liquidates the remaining $Q_\tau^v$ units for $S_\tau-\alpha Q_{\tau}^v$ per unit, where $\alpha$ represents the terminal liquidation penalty. 

The agents’ performance criteria is now defined as follows,
\begin{align}
H^{\nu}(t,S,q) = \mathbb{E}_{t,s,q}\displaystyle{\left[\int_t^{\tau}(S_u-f(\nu_u))du+Q_{\tau}(S_{\tau}-\alpha Q_{\tau})-\phi \int_t^{\tau}Q_u^2du\right]}
\label{eq:perf_crit_liq}
\end{align}
Here, $\phi \int_t^{\tau}Q_u^2du$ with $\phi \ge{0}$ is a running inventory penalty of the remaining units to be liquidated. The value function is next defined as
\begin{align}
H(t,S,q) = \sup_{\nu \in \mathscr{A}}H^{\nu}(t,S,q), \forall 0 \le t \le T, S_{min} \le S, 0 \le q \le \mathfrak{N}
\label{eq:val_func}
\end{align}
Here, $\mathscr{A}$ is again the set of admissible strategies in which $\nu >0$ and uniformly bounded from above. 

Next, we again set $g(\nu)=b\nu$ and $f(\nu)=\kappa\nu$ as in \cite{cartea2015algorithmic} and assume that our permanent and temporary impact functions are linear in the speed of trading, where $b \ge 0$, and $k>0$ are finite constants. Then, by applying the DPP, the value function should satisfy the following DPE,
\begin{align}
\begin{split}
&\partial_t H(t,S,q) + \frac{1}{2}(\sigma^2 +\bar{\sigma}^2 + \varsigma^2)\partial_{ss} H(t, S, q)-\phi q^2\\& + \sup_{\nu \in \mathcal{A}}\{-\nu\partial_qH(t,S,q)-b\nu\eta\partial_sH(t,S,q)+(S-\kappa\nu)\nu\}=0
\end{split}
\label{eq:DPE_liq}
\end{align}
subject to the terminal and boundary conditions:
\begin{align}
H(T,S,q)=(S-\alpha q)q;H(t,S_{min},q)=(S_{min}-\alpha q)q;H(t,S,0)=0
\label{eq:TC_BC_liq}
\end{align}
Here, remember the coefficients $\eta$, $\bar{\sigma}$ and $\varsigma$ are representing either the cases for $\eta_{SM}$, $\bar{\sigma}_{SM}$ and $\varsigma_{SM}$, or $\eta_{HP}$, $\bar{\sigma}_{HP}$ and $\varsigma_{HP}$, from Equations \eqref{eq:mp_SM_difapprox} and \eqref{eq:mp_HP_difapprox}, respectively. One can see from the terminal and boundary conditions, that once the stopping time is reached, the agent will liquidate the remainder of the units. Also, one can see that once the agent has liquidated the $\mathfrak{N}$ units initially targeted, the agent stops trading and there is no penalty, hence the boundary condition $H(t,S,0)=0$ where $q=0$. 

Then, as in Section 3.1, the optimal trading speed can be obtained by using the mathematical procedure called completing the squares on the inside of the minimization part of Equation \eqref{eq:DPE_liq}. And so, the optimal trading speed in feedback form can be defined as,  
\begin{align}
\nu^*(t,S,q) = -\frac{1}{2\kappa}(b\eta\partial_sH(t,S,q)+\partial_qH(t,S,q)-S).
\label{eq:control_liq}
\end{align}
One can then substitute this optimal control back into the DPE in Equation \eqref{eq:DPE_liq} and we then get the PDE:
\begin{align}
\begin{split}
&\partial_tH(t,S,q)+\frac{1}{2}(\sigma^2+\bar{\sigma}^2 + \varsigma^2)\partial_{SS}H(t,S,q)\\ &-\frac{1}{4\kappa}(b\eta\partial_sH(t,S,q)+\partial_qH(t,S,q)-S)^2 - \phi 	q^2 = 0
\end{split}
\label{eq:PDE_liq}
\end{align}
subject to the same terminal and boundary conditions defined in Equation \eqref{eq:TC_BC_liq}.

Following \cite{cartea2015algorithmic} and Section 3.1, we simplify the dimensionality to facilitate a numerical solution to this problem. To achieve this, we first set $b=0$ using the same argument as in Section 3.1 that the permanent impact of walking the LOB tends to be relatively small compared to the temporary impact. Then, we can use the following ansatz,
\begin{align}
H(t,S,q) = qS + q^2 h(t,S)
\label{eq:ansatz_liq}
\end{align}
and we see that $h$ satisfies the following PDE,
\begin{align}
\partial_t h(t,S)+\frac{1}{2}(\sigma^2 +\bar{\sigma}^2 + \varsigma^2)\partial_{SS}h(t,S)-\frac{1}{\kappa}h^2 - \phi = 0.
\label{eq:PDE_h_liq}
\end{align}
The existence of solutions to this PDE follows along similar lines as mentioned in Section 3.1 for the acquisition problem, where the general case for the Liquidation problem is also studied in \cite{cartea2015algorithmic}. This PDE now has a new terminal and boundary condition defined as,
\begin{align}
h(T,S) = \alpha,  S_{min} \le S,\label{eq:TC_h_liq}\\
h(t,S_{min}) = \alpha, t \le T.
\label{eq:BC_h_liq}
\end{align}
The optimal control for this liquidation problem, $\nu^*$, can now be defined as,
\begin{align}
\nu^*(t,S) = \frac{1}{\kappa}qh(t,S).
\label{eq:control_h_liq}
\end{align}
Intuitively, Equation \eqref{eq:control_h_liq} states that the optimal liquidation speed decreases as the inventory Q decreases, which would make sense as you are then closer to liquidating the total targeted units.  

Lastly, as in \cite{cartea2015algorithmic} for the acquisition problem, we would like to portray how the current status of the inventory process, $Q$, can be obtained in terms of the path of the midprice process by computing $Q_t$, the process that determines how much inventory is left to liquidate. Here, $Q_t$ satisfies the SDE,
\begin{align}
dQ_t^* = -\nu^*dt = -\frac{1}{\kappa}Q_th(t,S_t)dt,
\label{eq:inv_h_liq}
\end{align}
and therefore, 
\begin{align}
Q_t^* = \left(1-e^{ \frac{1}{\kappa}\int_0^th(u,S_u)du}\right)\mathfrak{N}, \ t \le \tau.
\label{eq:opt_inv_h_liq}
\end{align}
Here, we can see that the midprice path taken is paramount in computing the level of inventory the agent has liquidated at any given time.


\section{Numerical Solution: Acquisition and Liquidation}

In this Section, we will discuss the numerical solution to the acquisition and liquidation problems from Section 3.1 and 3.2, respectively. Here, we adopt an asset-specific approach, focusing on Microsoft (MSFT), with LOB data sourced from \cite{LOBSTER}. They offer free data on one day, June 21st 2012, on this asset. Building on this, we utilize the previously calibrated parameters from \cite{swishchuk2020general} and \cite{roldan2023stochastic}, assuming that Equation \eqref{eq:mp_nosubscript} follows the Hawkes dynamics described in Equation \eqref{eq:mp_HP_difapprox}. In addition, we conducted a sensitivity analysis to examine how varying these parameters affects the results. We solve the PDEs in Equation \eqref{eq:pde_h_acq} and \eqref{eq:PDE_h_liq} numerically using an Implicit-Explicit(IMEX) finite difference scheme, similarly to how they solve an acquisition problem in \cite{cartea2015algorithmic} for the pure diffusion case. The solutions are placed on a $[0,T]\times [S_{min}, S_{max}]$ grid, which is the domain. Then, for $n=0,1,...,N\in\mathbb{N}$, define $\Delta t=T/N$ such that $t_n=n\Delta t$ and for $i=0,1,...,M\in\mathbb{N}$, define $\Delta S=(S_{max}-S_{min})/M$ such that $S_i=S_{min}+i\Delta S$. The grids are divided into equally spaced nodes of distance $\Delta t$ and $\Delta S$, with mesh points $(n \Delta t,S_{min}+i \Delta S)$. We are concerned with the values of $h(t,S)$ i.e., the speed of trading per unit of inventory left to acquire/liquidate. The numerical scheme solves the $\partial_th$ and $\partial_{ss}h$ partial derivatives in Equation \eqref{eq:pde_h_acq} and \eqref{eq:PDE_h_liq} implicitly using standard methods and explicitly for the quadratic term as in \cite{cartea2015algorithmic}.

\subsection{Acquisition Problem Solution}
For the acquisition problem, we must specify a boundary condition along $S_{min}\ll S_{max}$ in order to have a well-posed problem. Here, we specify the boundary condition $\partial_{ss}h|_{S=S_{min}}=0$ as in \cite{cartea2015algorithmic}.  The numerical scheme, for $i = 1,2,..., M-1$ and $n = 0,1,..., N-1$, can then be derived as follows:
\begin{align}
\frac{h_i^{n+1}-h_i^n}{\Delta t}+\left(\frac{\sigma^2+\bar{\sigma}^2 +\varsigma^2}{2}\right)\frac{h_{i+1}^n+h_{i-1}^n-2h_i^n}{\Delta S^2}-\frac{1}{\kappa}\displaystyle\left(h_i^n\right)^2+\phi=0,
\label{eq:pde_fd_h_acq}
\end{align}
where $\bar{\sigma}^2$ and $\varsigma$ is either as defined for the Semi-Markov or Hawkes process cases in Section 2-3. Simplifying this leads to:
\begin{align}
h_i^{n+1}=-\alpha h_{i-1}^n+(1-\beta)h_i^n-\gamma h_{i+1}^n  + \frac{\Delta t}{\kappa}\displaystyle\left(h_i^n\right)- \Delta t\phi
\label{eq:pde_fd_simplified_h_acq}
\end{align}
where $\alpha = \gamma = (\sigma^2+\bar{\sigma}^2+\varsigma^2)\frac{\Delta t}{2\Delta S^2}$ and $\beta=(\sigma^2+\bar{\sigma}^2+\varsigma^2)\frac{\Delta t}{\Delta S^2}$. Then, partially in matrix form, the solution, again at $i = 1,2,..., M-1$ and $n = 0,1,..., N-1$, can be calculated as:
\begin{align}
h^n=M_1^{-1}\displaystyle\left(h^{n+1} -\frac{\Delta t}{\kappa}(h_i^n)^2+\Delta t \phi\right)
\label{eq:h_solution_acq}
\end{align}
where $M_1$ is defined as,
\[ M_1=\begin{bmatrix}
   1-\beta_1 & -\gamma_1  &  & \\ 
   -\alpha_2 & 1-\beta_2 &-\gamma_2  & \\ 
   &  -\alpha_3 & 1-\beta_3 &-\gamma_3 \\
   &  \ddots&  \ddots &\ddots& \\
   &  & -\alpha_{M-2} & 1-\beta_{M-2} & -\gamma_{M-2}\\ 
   &  &  & -\alpha_{M-1}& 1-\beta_{M-1} 
\end{bmatrix} \]
Note that one must also account for the boundary conditions, at $i = 0$ and $i = M$, and the terminal condition, at $n = N$. These were defined in Equation \eqref{eq:TC_h_Acq}-\eqref{eq:BC_h_Acq} and the condition above where a lower boundary was set at $S_{min}$.

\begin{figure}[ht]
	\centering
	\includegraphics[width=\linewidth]{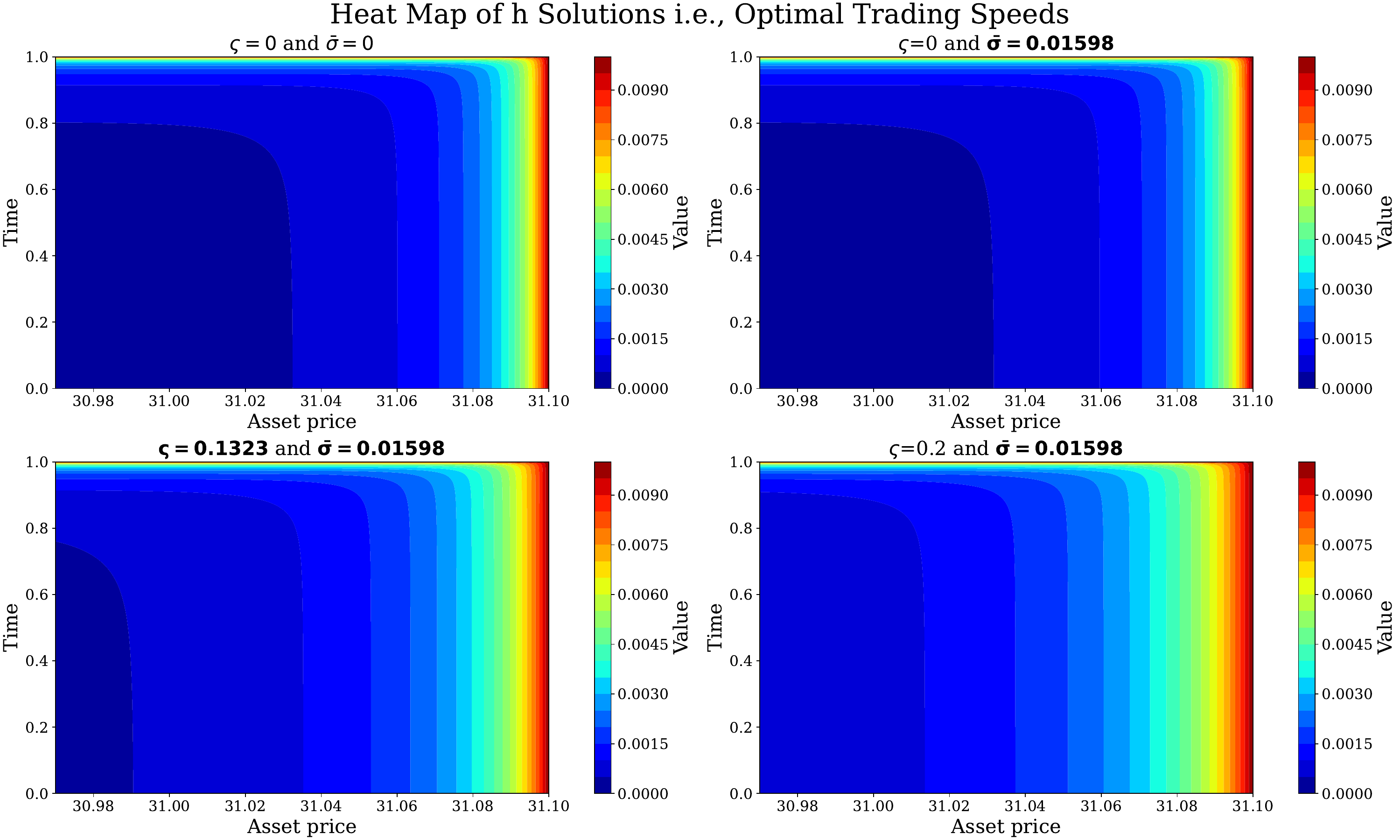}
	\caption{Optimal trading speed as the time and asset price increases for the acquisition problem.}
	\label{fig:h_solutions}
\end{figure}

In Figure \ref{fig:h_solutions}, we show a set of subplots which display contour plots portraying how the solution changes as a function of time and the financial asset's price, over different values of $\bar{\sigma}$ and $\varsigma$, where the calibrated parameter values are indicated in bold. Time here is from $t=0$ to the maturity time $t=T=1$, and the midprice (the space variable) is from $S_0=30.97$ to $S_{max}=31.1$. We specifically focus on this part of the solution as this is where it varies the most. Right of each subplot shows the color scheme for the values of the solution that are displayed. These solution values vary from $0$ to $0.01$, where $\alpha=0.01$ represents the value of the terminal condition and the boundary condition at $T$ and $S_{max}$, respectively, as defined in Equations \eqref{eq:TC_h_Acq}-\eqref{eq:BC_h_Acq}.  We can see that the agents speed of acquiring units increases as we get closer to the terminal time. This is because the agent is trying to avoid the terminal penalty cost that they must pay to purchase any remaining units at maturity. Furthermore, the speed of trading also increases as the asset's price approaches the limit price, $S_{max}$, again to avoid the other terminal penalty. This trading process was formalized in the stopping time in Equation \eqref{eq:stop_time_Acq}.   

In the top left subplot of Figure \ref{fig:h_solutions}, we show the solution for the case where both $\bar{\sigma}=0$ and $\varsigma=0$, which would represent the case for benchmark models in \cite{cartea2015algorithmic}. In other words, this subplot would portray the same solution were our price process to follow the pure diffusion dynamics given in Equation \eqref{eq: mp_cartea}. In the top right subplot, we show the solution for the case where $\bar{\sigma}=0.01598$ as calibrated and $\varsigma=0$, which shows the same solution if the pricing model follows the price process from \cite{cartea2015algorithmic} combined with the price processes developed in \cite{roldan2022optimal} and \cite{roldan2023stochastic}. Recall that the latter models approximate the price process, via a diffusion approximation, for the cases where the dynamics approximate the semi-Markov or Hawkes processes. The two bottom subplots, where we increased the value of the parameter $\varsigma$ to its calibrated parameter value and then a higher value, show the solutions for the cases where the price process follows our more general jump-diffusion model given in Equation \eqref{eq:mp_nosubscript}. 

In each consecutive subplot, from left to right, we vary either $\bar{\sigma}$ or $\varsigma$ to show the effect that the Semi-Markov or Hawkes process parts could have on the optimal solution. One can see from the subplots that as we slowly increment either of these parameters, the speed of trading increases at earlier time steps at lower asset prices. This makes sense intuitively, as the higher either of these coefficients becomes, the more overall volatility the solution predicts. Thus, in these scenarios, it would become more optimal to acquire units of the asset faster on the basis of our model. In this particular type of trading problem, where the agent is acquiring units, it is mainly price jumps that go up which can significantly change the optimal solution as the agent will want to acquire units of the asset a lot quicker then. In the case where the asset price would jump down, this would be highly favorable to the agent, and their speed of acquiring any remaining units of the asset would then mostly depend on time. 

To see the remaining input parameters for this model, please see Table \ref{tab:table1} below. Most of these parameters were kept similar to the simulation results in \cite{cartea2015algorithmic} for comparison purposes, where they provide some details of their choices. However, we did make some changes. Firstly, we increased the spatial size of the grid to better illustrate how our new parameters can influence a problem of this nature. To do this, we slightly increase the distance between $S_0$ and $S_{max}$ and decrease $S_{min}$ by a greater amount as this part of the solution is less relevant as lower prices reduce the problem to a time-dependent execution strategy. Secondly, we decreased the value of $\phi$ from $0.001$ to $0.00001$, the running inventory penalty, so that our solution mostly focuses on the effect of the Semi-Markov and Hawkes process parts of the model. This parameter must be nonzero for our strategy simulations later which is why we decided to make it very small instead of setting it to zero. Next, as explained in \cite{cartea2015algorithmic}, the formula $\alpha/\kappa$ represents the maximum trading speed, which is why $\alpha=0.01$. The choice of volatility, which was calibrated independently on the MSFT data using the maximum likelihood estimator approach, was set to $\sigma = 0.1041$. These are also mostly the same parameter values used for our strategy simulations and similar to the ones that \cite{cartea2015algorithmic} used in their strategy simulations, which we will show in Section 5. 

\begin{table}[H]
    \centering
    \small 
    \begin{tabular}{ |c|c|c|c| }
    \hline
    \multicolumn{4}{|c|}{\textbf{Parameters}} \\
    \hline
    \textbf{Parameter} & \textbf{Value} & \textbf{Parameter} & \textbf{Value} \\
    \hline
    $\bar{\sigma}$ & $0.01598$ & $\varsigma$ & $0.1323$ \\
    $\sigma$ & 0.1041 & $S_0$ & 30.97 \\
    T & 1 & $\alpha$ & 0.01 \\
    $\kappa$ & 0.0001 & $\phi$ & 0.00001 \\
    $N$ & 390 & $M$ & 1000 \\
    $S_{min}$ & 29 & $S_{max}$ & 31.1 \\
    $\Delta t$ & 0.002 & $\Delta S$ & 0.002 \\
    \hline
    \end{tabular}
    \caption{Simulation parameters.}
    \label{tab:table1}
\end{table}

\subsection{Liquidation Problem Solution}

For the liquidation problem, as for the acquisition problem in Section 4.1, we must ensure that we have a well-posed problem. To achieve this, we now specify a boundary condition along $S_{max}\gg S_{min}$, similarly to how a boundary condition was specified along $S_{min}\ll S_{max}$ in Section 4.1. Here, we specify the boundary condition $\partial_{ss}h|_{S=S_{max}}=0$. Then, the numerical scheme, again for $i = 1,2,..., M-1$ and $n = 0,1,..., N-1$, for the liquidation problem, where its PDE was defined in Equation \eqref{eq:PDE_h_liq}, can be derived as follows:
\begin{align}
\frac{h_i^{n+1}-h_i^n}{\Delta t}+\left(\frac{\sigma^2+\bar{\sigma}^2+\varsigma^2}{2}\right)\frac{h_{i+1}^n+h_{i-1}^n-2h_i^n}{\Delta S^2}-\frac{1}{\kappa}\displaystyle\left(h_i^n\right)^2-\phi=0,
\label{eq:pde_fd_liq}
\end{align}
where $\bar{\sigma}$ and $\varsigma$ is again either as defined for the Semi-Markov or Hawkes process cases given in Sections 2-3. Simplifying this leads to:
\begin{align}
h_i^{n+1}=-\alpha h_{i-1}^n+(1-\beta)h_i^n-\gamma h_{i+1}^n  + \frac{\Delta t}{\kappa}\displaystyle\left(h_i^n\right)+ \Delta t\phi
\label{eq:pde_fd_simplified_h_liq}
\end{align}
where again $\alpha = \gamma = (\sigma^2+\bar{\sigma}^2+\varsigma^2)\frac{\Delta t}{2\Delta S^2}$ and $\beta=(\sigma^2+\bar{\sigma}^2+\varsigma^2)\frac{\Delta t}{\Delta S^2}$, which is the same as in Section 4.1. Then, partially in matrix form, the solution to the numerical scheme, again at $i = 1,2,..., M-1$ and $n = 0,1,..., N-1$, can be calculated as:
\begin{align}
h^n=M_1^{-1}\displaystyle\left(h^{n+1} -\frac{\Delta t}{\kappa}(h_i^n)^2-\Delta t \phi\right)
\label{eq:h_solution_liq}
\end{align}
where $M_1$ is defined as in Section 4.1. Note that here one must also account for the boundary conditions at $i = 0$, $i = M$ and for the terminal condition at $n = N$. These are as defined in Equations \eqref{eq:TC_h_liq}-\eqref{eq:BC_h_liq} and the condition above where an upper boundary condition was set at $S_{max}$. One can see that Equation \eqref{eq:h_solution_liq} is very similar to Equation \eqref{eq:h_solution_acq}, which was used to solve the acquisition problem. The main changes here are in the boundary conditions. The liquidation problem now has a price floor rather than a price cap, as in the acquisition problem. 

\begin{figure}[H]
	\centering
	\includegraphics[width=\linewidth]{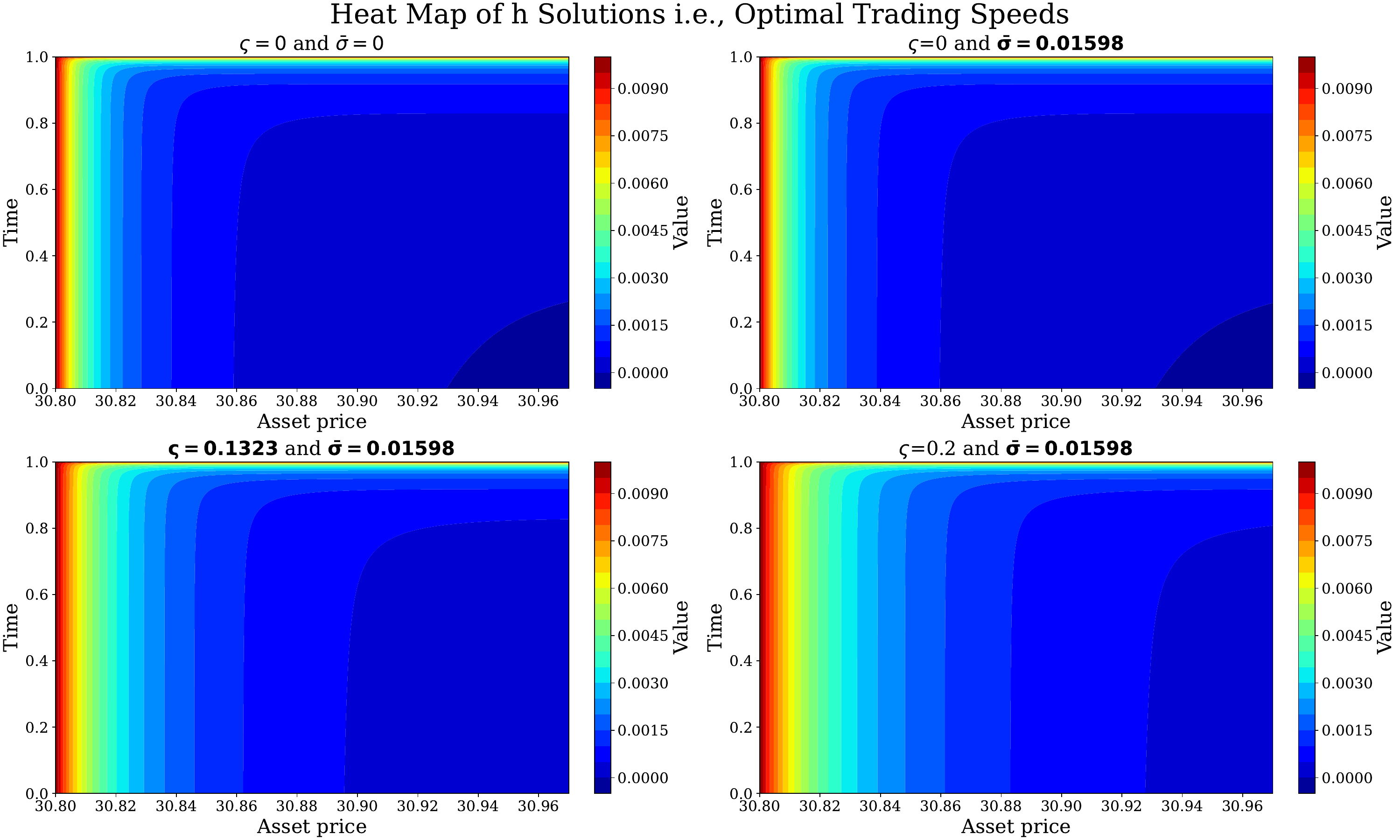}
	\caption{Optimal trading speed as the time and asset price decreases for the liquidation problem.}
	\label{fig:h_solutions_liq}
\end{figure}

Next, in Figure \ref{fig:h_solutions_liq}, we show a set of subplots which display contour plots portraying how the solution changes as a function of time and the financial asset's price, over different values of $\bar{\sigma}$ and $\varsigma$. Time, here, is again from $t=0$ to the maturity time $t=T=1$, and the midprice (the space variable) is from $S_{min}=30.8$ to $S_0=30.97$. We specifically focus on this part of the solution as this is where it varies the most. Right of each subplot shows the color scheme for the values of the solution that are displayed. These solution values vary from $0$ to $0.01$, where $\alpha=0.01$ represents the value of the terminal condition and the boundary condition at $T$ and $S_{min}$, respectively, as defined in Equations \eqref{eq:TC_h_liq}-\eqref{eq:BC_h_liq}. We can see that the agents speed of liquidating units increases as we get closer to the terminal time. This is because the agent is trying to avoid the terminal penalty cost that they must pay to liquidate any remaining units at maturity. Furthermore, the speed of trading also increases as the asset's price approaches the floor price, $S_{min}$, again to avoid the other terminal penalty. This trading process was formalized in the stopping time in Equation \eqref{eq:stop_time_liq}.   

As in Figure \ref{fig:h_solutions}, the format for the pricing models in these plots is similar. In the top left subplot of Figure \ref{fig:h_solutions_liq}, we show the solution to the liquidation problem for the case where both $\bar{\sigma}=0$ and $\varsigma=0$, which would represent the case for the pure diffusion benchmark models in \cite{cartea2015algorithmic}. In the top right subplot, we show the solution for the case where $\bar{\sigma}=0.01598$ and $\varsigma=0$, which would portray the same solution were the pricing model to follow the price process in \cite{cartea2015algorithmic} combined with the price processes developed in \cite{roldan2022optimal} and \cite{roldan2023stochastic}. The two bottom subplots, where we increased the value of the parameter $\varsigma$, where the bold value represents the calibrated parameter value, then show the solutions for the cases where the price process follows our more general jump-diffusion model given in Equation \eqref{eq:mp_nosubscript}. 

In each consecutive subplot, from left to right, we again vary either $\bar{\sigma}$ or $\varsigma$ to show the effect that the Semi-Markov and Hawkes process parts can have on the optimal solution. In each of these subplots, one can see the effect on the solution matrix as we slowly incremented the effect of either of the parameters, $\bar{\sigma}$ or $\varsigma$. We can see that as they increase, the speed of trading increases at earlier time steps at lower asset prices. This makes sense intuitively, as the higher these coefficients become, the more overall volatility the solution predicts. Thus, in these scenarios, it would then be more optimal to liquidate units of the asset quicker based on our model. In this particular type of trading problem, where the agent is liquidating units, it's mainly price jumps that go down which can significantly change the optimal solution as the agent will want to liquidate units of the asset a lot quicker then. In the case where the asset price would jump up, this would be highly favorable to the agent, and the optimal speed of liquidating any remaining units of the asset would then mostly depend on time.  

Most of the input parameters for this model are kept the same as in Table \ref{tab:table1} in Section 4.1 for the acquisition problem. The only value that we changed was $S_{min}$ and $S_{max}$, which here were set to $30.8$ and $33$, respectively. These were changed so that we could focus more on the impact of the price floor on the optimal solution, whereas in Section 4.1 the solution is more focused on the effect of the price cap. 

\section{Strategy Simulations: Acquisition and Liquidation}

In this Section, we simulate the performance of the above acquisition and liquidation problems over $10,000$ different price paths, which is the total number of simulations we ran. In Section 4, the optimal solution takes a static view of the markets, whereas in reality, market conditions are constantly evolving. This should be reflected in the agent's strategy and is throughout our strategy simulations, where the strategy is updated at each time step to reflect this. Here, we will, again, specifically focus on how an increasing $\bar{\sigma}$ or $\varsigma$ can significantly change how the acquisition/liquidation strategy would evolve and how the certain processes we defined in Section 3 are affected. To setup these simulations, one must discretize the continuous-time processes introduced in Section 3. In other words, the continuous-time processes in Equations \eqref{eq:inv_process}, \eqref{eq:mp_nosubscript}, \eqref{eq:exec_price}, \eqref{eq:cash_process}, \eqref{eq:remaining_inv_acq}, \eqref{eq:control_h_acq}-\eqref{eq:opt_inv_h_acq} and \eqref{eq:control_h_liq}-\eqref{eq:opt_inv_h_liq} are discretized. Since both the acquisition and liquidation trading problems have already been solved numerically, as shown in Section 4, this part is already given to us in a discrete form, and we can use the values in these solution matrices to guide our trading decisions in the strategy simulation. As we increment forward in time discretely, each simulation looks at the current price at every time step and then selects the matching value from the $h$ solution matrix, which guides the agent on what the new optimal trading speed is. After a trade is made, the cash, inventory, and execution price processes are updated. If either the terminal or boundary condition is breached, the simulation ends and the agent pays the terminal penalty to acquire/liquidate any remaining units of the asset. In this Section we will again present subplots in the same format as in Figure 1 and 2. Recall that in those Figures, that the top left subplot is for the \cite{cartea2015algorithmic} case, the top right subplot is for the \cite{cartea2015algorithmic} case combined with the cases in \cite{roldan2022optimal} and \cite{roldan2023stochastic}, and the bottom two subplots are for our new more general price processes as defined in Equation \eqref{eq:mp_nosubscript}.

\subsection{Acquisition Problem Strategy Simulations}

In this Section we will discuss the strategy simulations for the acquisition problem. These are very similar to the simulations performed in \cite{cartea2015algorithmic} in their pure diffusion models. First, see below, in Figure \ref{fig:price_paths}, the first five sample price paths from our simulations and, in Figure \ref{fig:price_histograms}, a histogram showing the frequency of the average traded price in each of the $10,000$ simulations. In Figure \ref{fig:price_paths}, these 5 random price paths are based on our price process defined earlier in Equation \eqref{eq:mp_nosubscript}, where we repeat the simulations for different values of $\bar{\sigma}$ or $\varsigma$. In Figure \ref{fig:price_histograms}, one can see how the different parameter values can influence the asset price the agent ends up acquiring units at in our strategy simulations. The dashed black line in both Figures also shows the price cap, $S_{max}$, which if breached ends the trading strategy simulation. 

\begin{figure}[H]
	\centering
	\includegraphics[width=0.95\linewidth]{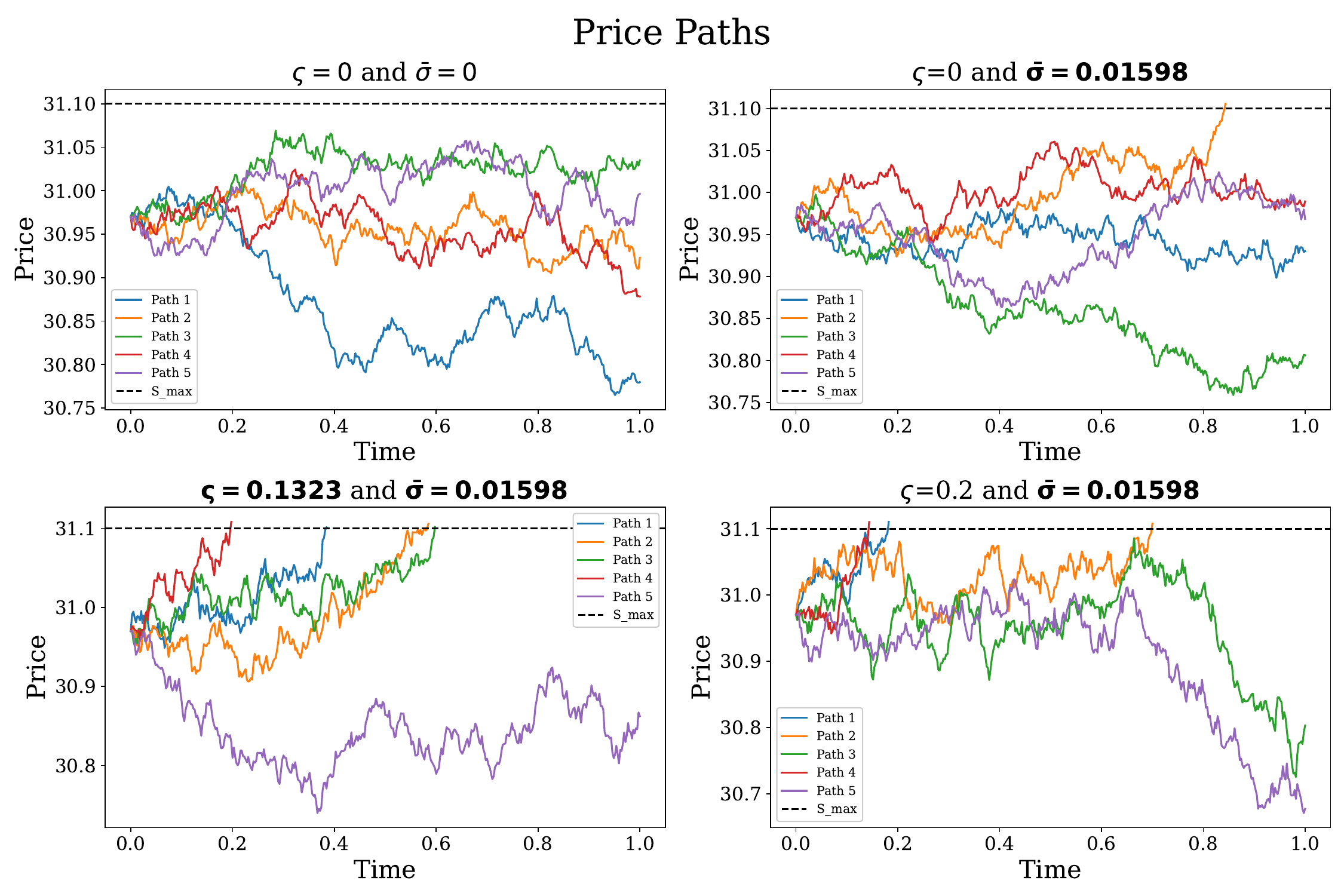}
	\caption{Five sample price paths over varying values of either $\bar{\sigma}$ or $\varsigma$.}
	\label{fig:price_paths}
\end{figure}

It is quite obvious that as the $\bar{\sigma}$ or $\varsigma$ values are increment, prices are more likely to exhibit higher volatility, as this is still a diffusion model, but we will see next how this higher volatility coming from the jump part (via diffusion approximation) can significantly alter how the simulated trading strategy will evolve. One thing to notice from these plots is that when our boundary condition $h(t, S_{max}=31.1)$ is breached. Here, we encounter one of the times where the agent purchases the remaining units instantly, pays the terminal penalty $\alpha = 0.01$ and stops trading. The other time this can occur is when $t=T=1$. We will notice in the next set of plots how the strategy is then significantly altered each time this occurs and we will portray how this can significantly effect the overall performance of the strategy in these scenarios. 

\begin{figure}[H]
	\centering
	\includegraphics[width=0.95\linewidth]{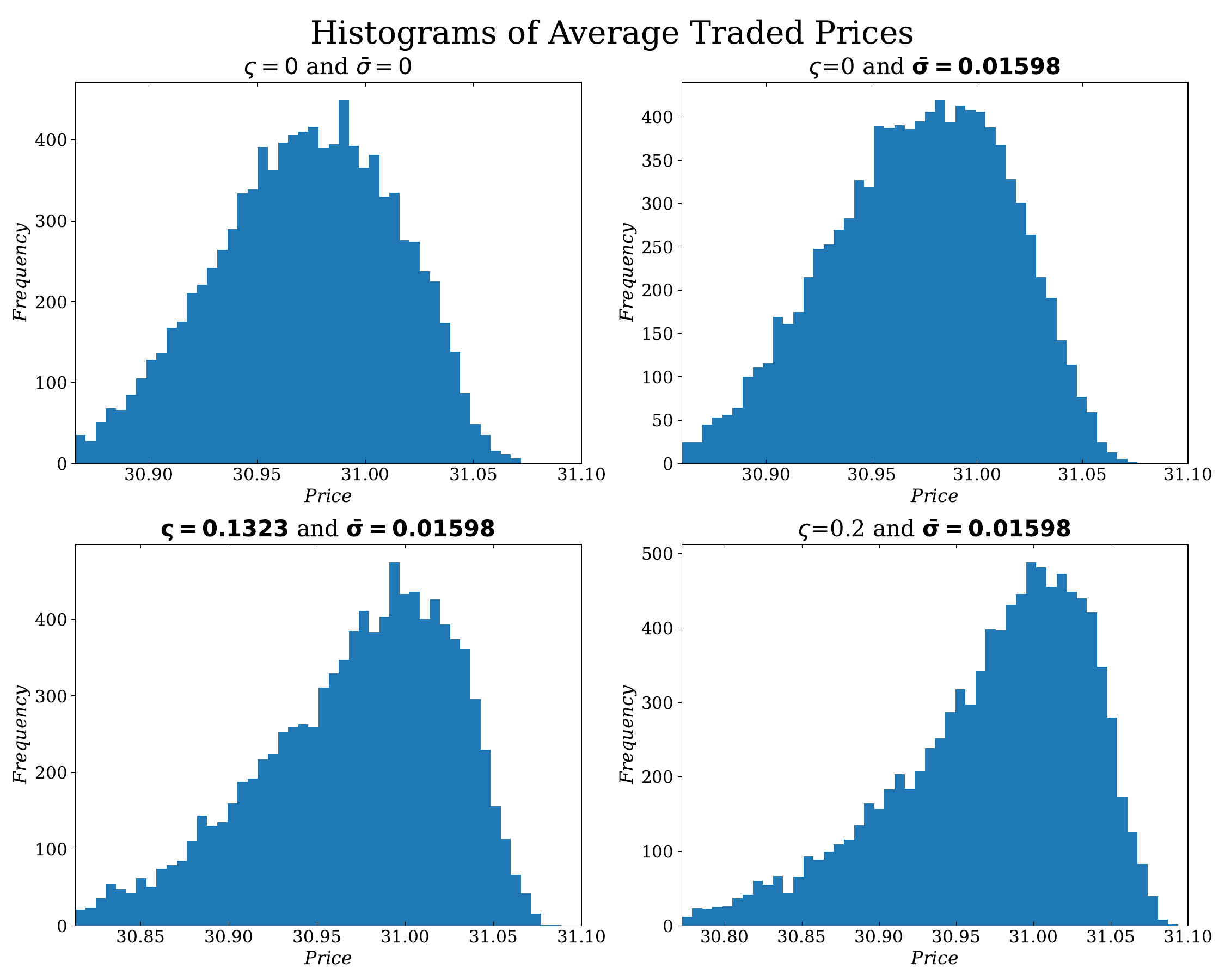}
	\caption{The average acquisition prices over all 10,000 simulations for varying values of either $\bar{\sigma}$ or $\varsigma$.}
	\label{fig:price_histograms}
\end{figure}

\begin{figure}[H]
	\centering
	\includegraphics[width=0.95\linewidth]{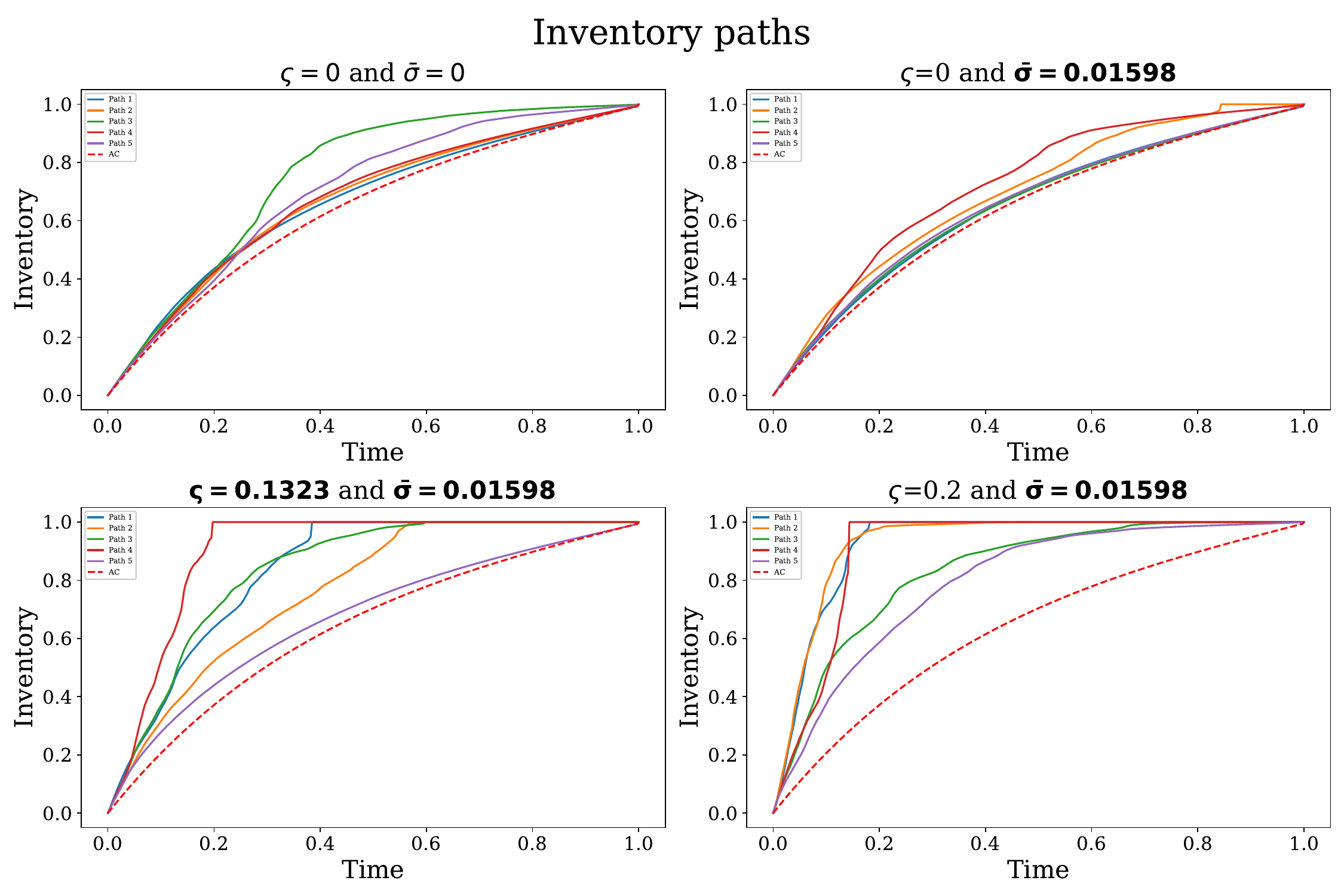}
	\caption{The inventory paths for the five sample price paths over varying values of either $\bar{\sigma}$ or $\varsigma$.}
	\label{fig:inventory_paths}
\end{figure}

\begin{figure}[H]
	\centering
	\includegraphics[width=0.95\linewidth]{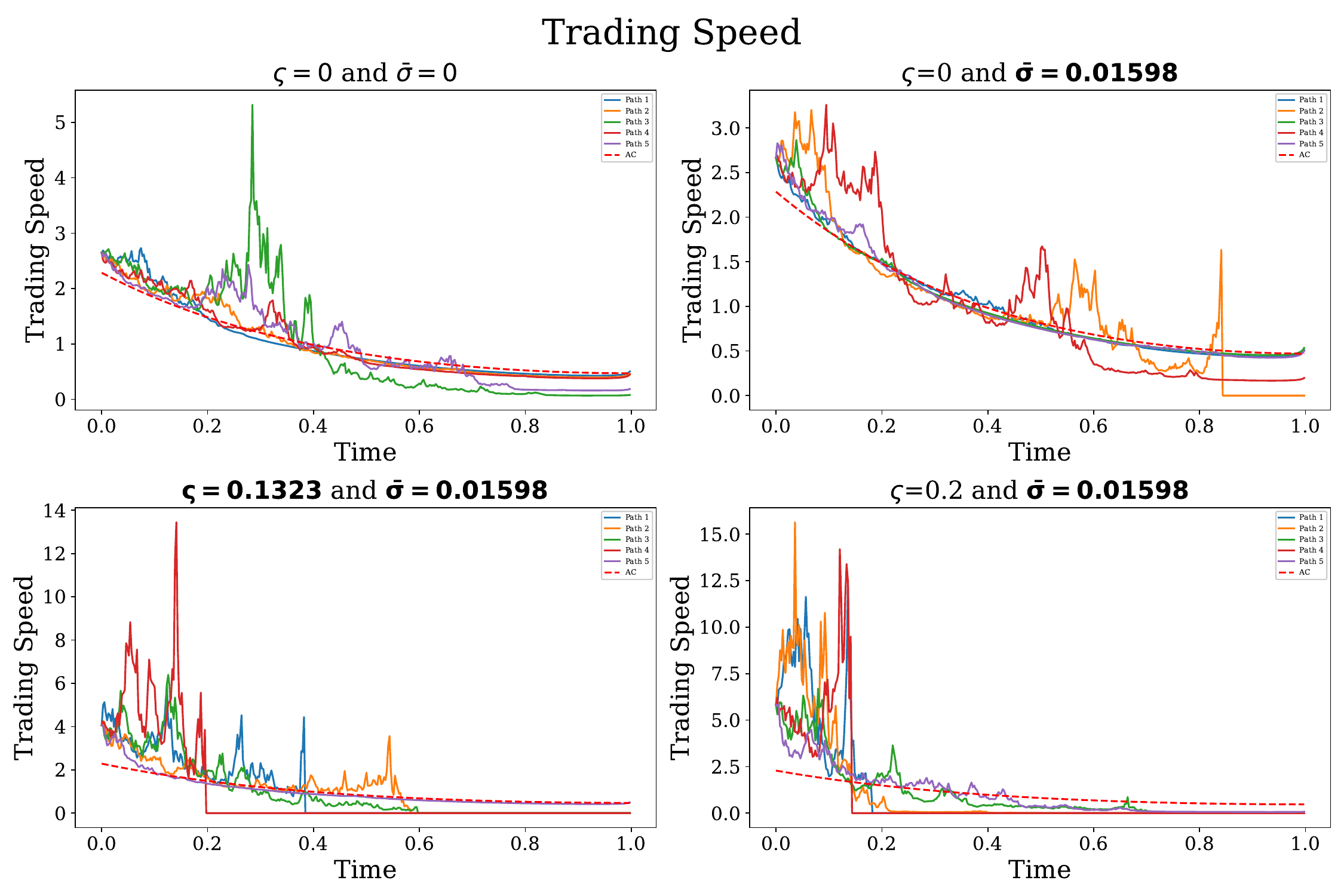}
	\caption{The trading speed paths for the five sample price paths over varying values of either $\bar{\sigma}$ or $\varsigma$.}
	\label{fig:trading_speed}
\end{figure}

Next, in Figures \ref{fig:inventory_paths} and \ref{fig:trading_speed}, we show how the agent's inventory and trading speed processes evolve over the five sample price paths we showed above in Figure \ref{fig:price_paths}. These are the discretized versions of the continuous-time processes given in Equations \eqref{eq:control_h_acq} and \eqref{eq:remaining_inv_h_acq}. In these Figures, we also plot the Almgren-Chriss (AC) strategy from \cite{almgren2001optimal}, which is the same benchmark used by \cite{cartea2015algorithmic} and part of the motivation for their models.  The AC strategy is deterministic and for the acquistion problem it can be defined, as given in \cite{cartea2015algorithmic}, by the following formula, 
\begin{align}
\nu_t^{AC} = \sqrt{\kappa\phi}\frac{\xi e ^{2\gamma(T-t)}+1}{\xi e ^{2\gamma(T-t)}-1},
\label{eq:trading_speed_AC_acq}
\end{align}
where $\xi = \frac{\alpha + \sqrt{\kappa\phi}}{\alpha - \sqrt{\kappa\phi}}$ and $\gamma = \sqrt{\frac{\phi}{\kappa}}$. This holds as long as $\phi >0$, which we defined earlier as the running inventory penalty. The AC strategy is a widely recognized benchmark for dealing with large sized trade orders, such as for the acquisition problem we studied. As it is a linear impact model, it is unchanged in all the subplots and is, thus, a good benchmark for assessing the performance of the optimal acquisition strategy.

The common theme throughout these subplots is that as we increment $\bar{\sigma}$ or $\varsigma$, these processes themselves become a lot more volatile and are heavily determined by the direction and the degree to which the price is moving along each path. We notice that in the top left subplot, the trading strategy (viewed through the lens of the inventory and trading speed processes) does not vary much along the different price paths from Figure \ref{fig:price_paths}. But as we introduce nonzero increasing values for either of the coefficients $\bar{\sigma}$ or $\varsigma$ from our models, a large portion of the simulations end up acquiring units of the asset a lot quicker and many have acquired the targeted inventory well before the maturity time, $T$. This is not too surprising, as the risk of obtaining good acquisition prices increases as these parameter values are incremented, thus the agent is induced to trade a lot quicker as a result of this added risk. This is quite important because if a trader were to run an acquisition strategy like this in live markets, and the price were to suddenly jump up, they would want the algorithm to react accordingly. In other words, if a sudden jump up were to enter the market, $\varsigma$ would instantly increase, inducing the trader to increase the speed of acquiring units before the maximum price limit is breached. This increase in the speed of trading would be higher the closer the agent is to the maturity time, $T$. Remember, that if prices were to jump down, this would actually be beneficial to the trader, as they would then end up buying at lower prices and there should be no increase in urgency to buy units of the asset quickly. Thus, we don't focus much on this case as the optimal trading speed is then mostly based on the time left to maturity, which would be quite similar to the AC model for the same number of remaining units of the asset left to purchase.

\begin{figure}[H]
	\centering
	\includegraphics[width=0.85\linewidth]{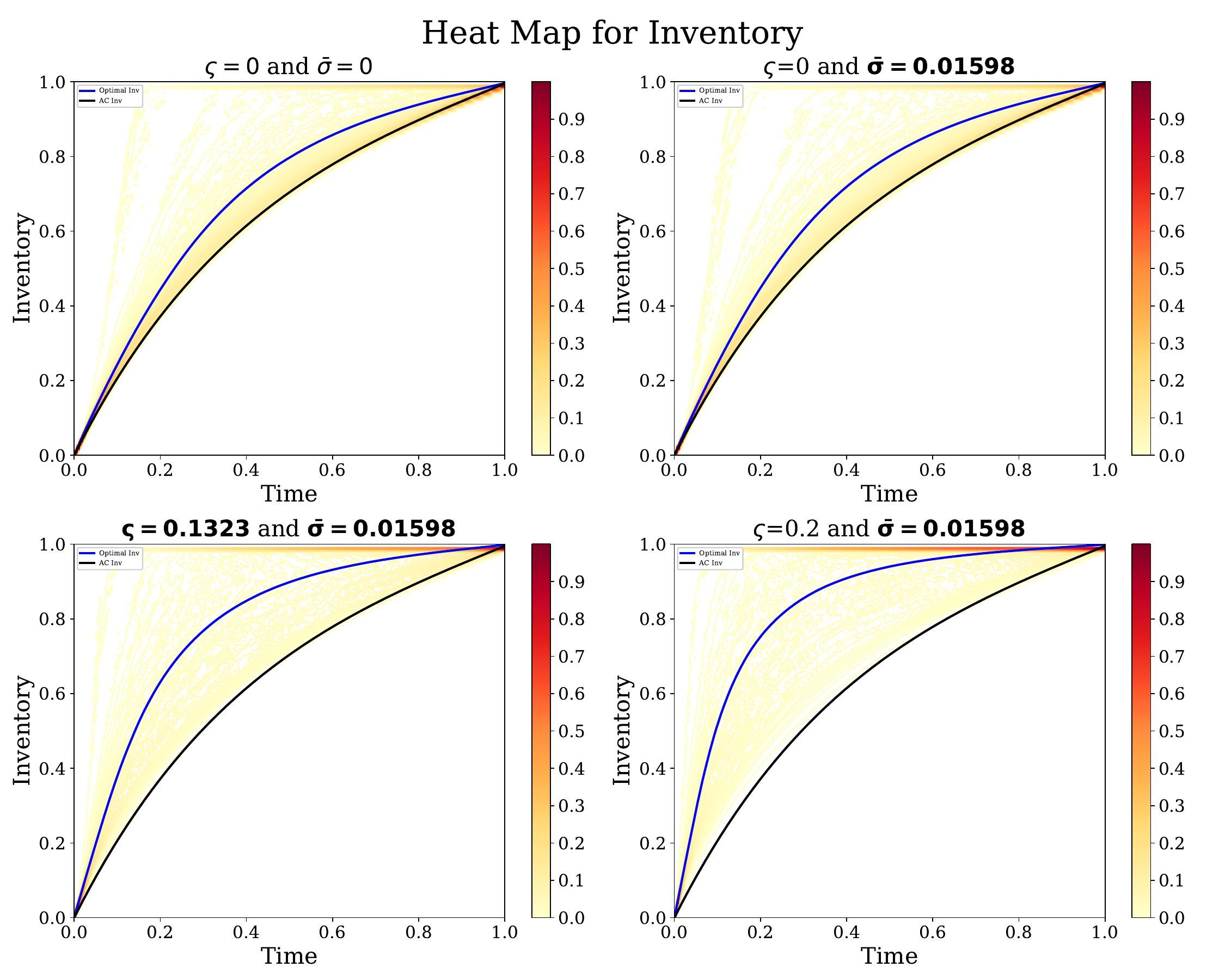}
	\caption{A heat map for the inventory paths over all 10,000 sample price paths over varying values of either $\bar{\sigma}$ or $\varsigma$.}
	\label{fig:inventory_heat_map}
\end{figure}

\begin{figure}[H]
	\centering
	\includegraphics[width=0.85\linewidth]{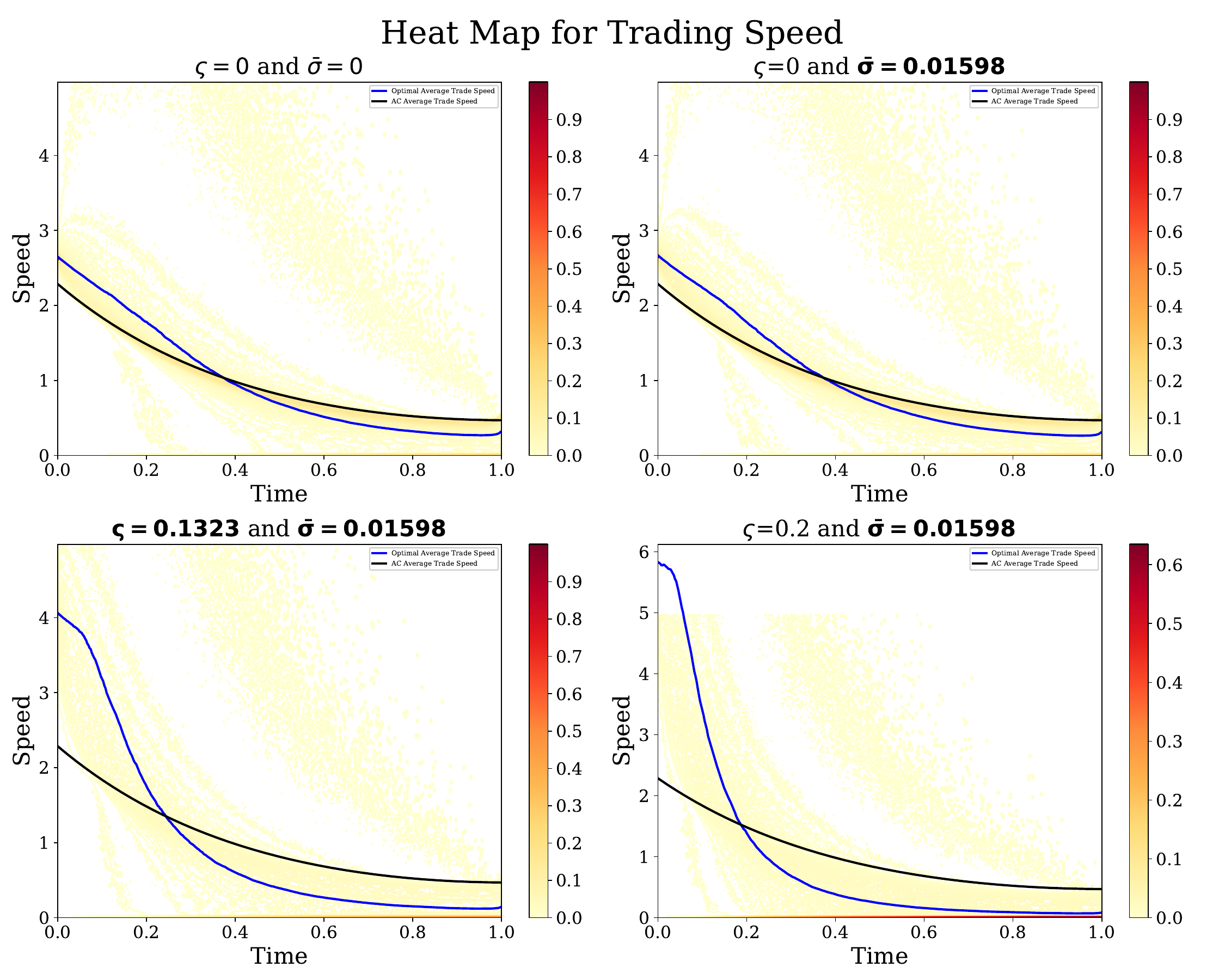}
	\caption{A heat map for the trading speed paths over all 10,000 sample price paths over varying values of either $\bar{\sigma}$ or $\varsigma$.}
	\label{fig:trading_speed_heat_map}
\end{figure}

Lastly, in Figures \ref{fig:inventory_heat_map} and \ref{fig:trading_speed_heat_map}, we show heat maps for the inventory and trading speed processes over all 10,000 simulations. The blue line in both graphs shows the mean of all the inventory and trading speed paths under our optimal strategy, while the black line shows the mean of the inventory and trading speed paths under the AC strategy, which recall is a benchmark strategy. In Figure \ref{fig:inventory_heat_map}, we notice that as we increment either of the parameters $\bar{\sigma}$ and $\varsigma$, the optimal inventory line becomes more concave shaped and thus moves away from the AC line, with this effect increasing the further we are from $t=0$ and $t=T$. In Figure \ref{fig:trading_speed_heat_map}, one will notice that as we increment either of the parameters $\bar{\sigma}$ and $\varsigma$, the trading speed will increase significantly earlier on in the simulation, and thus decreases later on as it acquired most of its targeted inventory a lot earlier. In both plots, we can see that there are a lot of points at the top of inventory plot (Figure \ref{fig:inventory_heat_map}) and the bottom of the trading speed plot (Figure \ref{fig:trading_speed_heat_map}). These are the simulations where the barrier at $S_{max}$ or $y=0$ got breached before $t=T$. There are more of these instances as we increment either of the parameter values $\bar{\sigma}$ and $\varsigma$, which would make sense intuitively, since when the overall level of volatility is higher, the agent is induced to complete the acquisition strategy a lot quicker. 

\subsection{Liquidation Problem Strategy Simulations}

Here, as in Section 5.1, we analyze how the strategy performed in our simulations but for the liquidation problem. First, see below, in Figure \ref{fig:price_paths_liq}, the first five sample price paths from our simulations and, in Figure \ref{fig:price_histograms_liq}, a histogram showing the frequency of the average traded price in each of the $10,000$ simulations. In Figure \ref{fig:price_paths_liq}, these 5 random price paths are again based on our price process defined earlier in Equation \eqref{eq:mp_nosubscript}, where we repeat the simulations for different values of either $\bar{\sigma}$ or $\varsigma$. The dashed black line in both plots now shows the price floor, $S_{min}$, which if breached ends the simulation. As in Section 5.1, Figure \ref{fig:price_histograms_liq} portrays how the price paths end up influencing the average asset prices the agent ends up liquidating units of the asset at.  It is, again, quite obvious that as we increment either $\bar{\sigma}$ or $\varsigma$, prices are more likely to exhibit higher volatility, as this is still a diffusion model, but we will see next how this higher volatility coming from the jump part (via diffusion approximation) can significantly alter how the simulated trading strategy will evolve. One thing to notice from these plots is that when our boundary condition $h(t, S_{min}=19.8)$ is breached. Here, we encounter one of the times where the agent liquidates the remaining units instantly, pays the terminal penalty $\alpha = 0.01$ and stops trading. The other time this would occur at is when $t=T=1$, like in the acquisition problem. We will notice in the next set of plots how the strategy is then significantly altered each time this occurs and we will portray how this can significantly effect the overall performance of the strategy in these scenarios. 

\begin{figure}[H]
	\centering
	\includegraphics[width=0.95\linewidth]{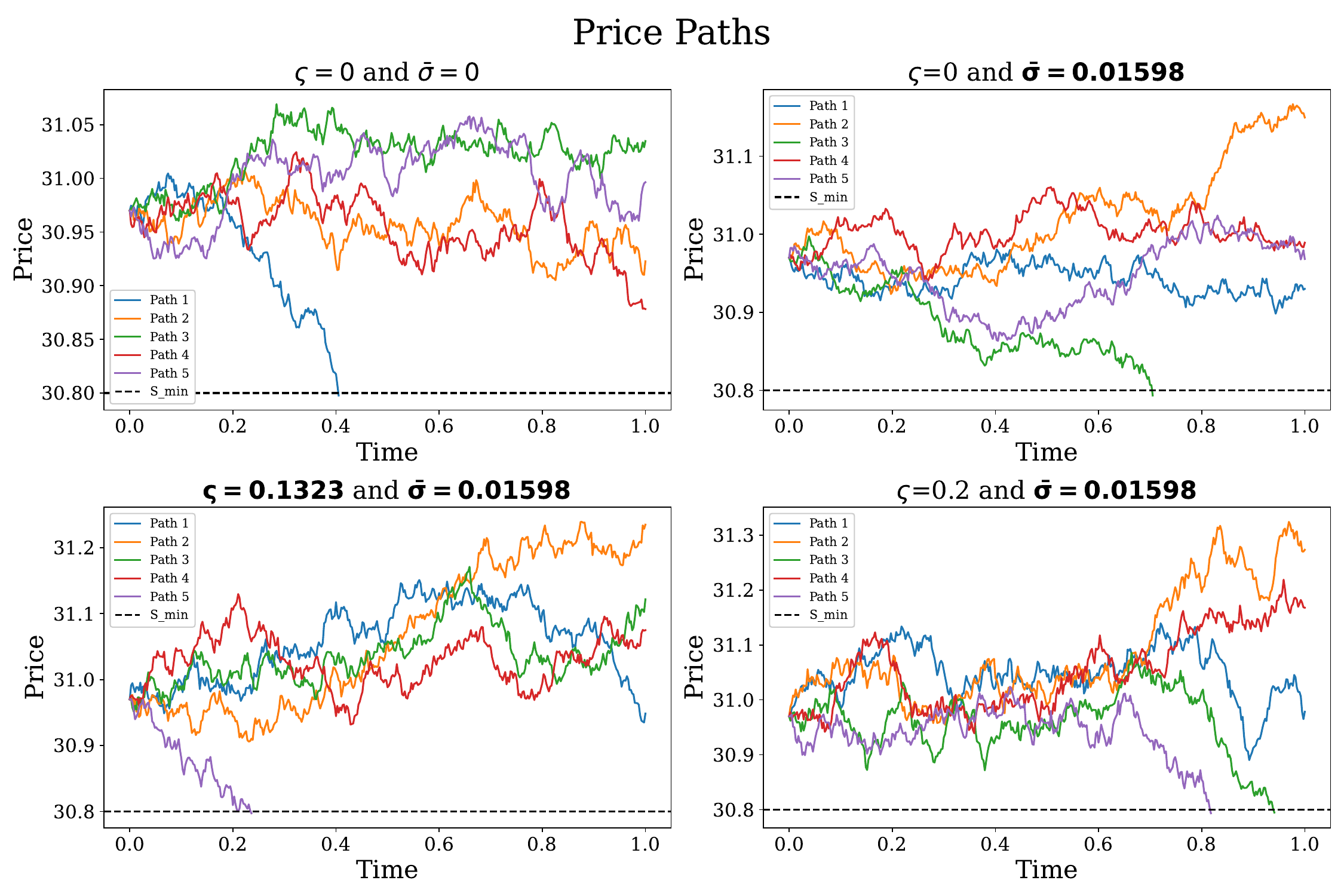}
	\caption{Five sample price paths over varying values of either $\bar{\sigma}$ or $\varsigma$.}
	\label{fig:price_paths_liq}
\end{figure}

\begin{figure}[H]
	\centering
	\includegraphics[width=0.95\linewidth]{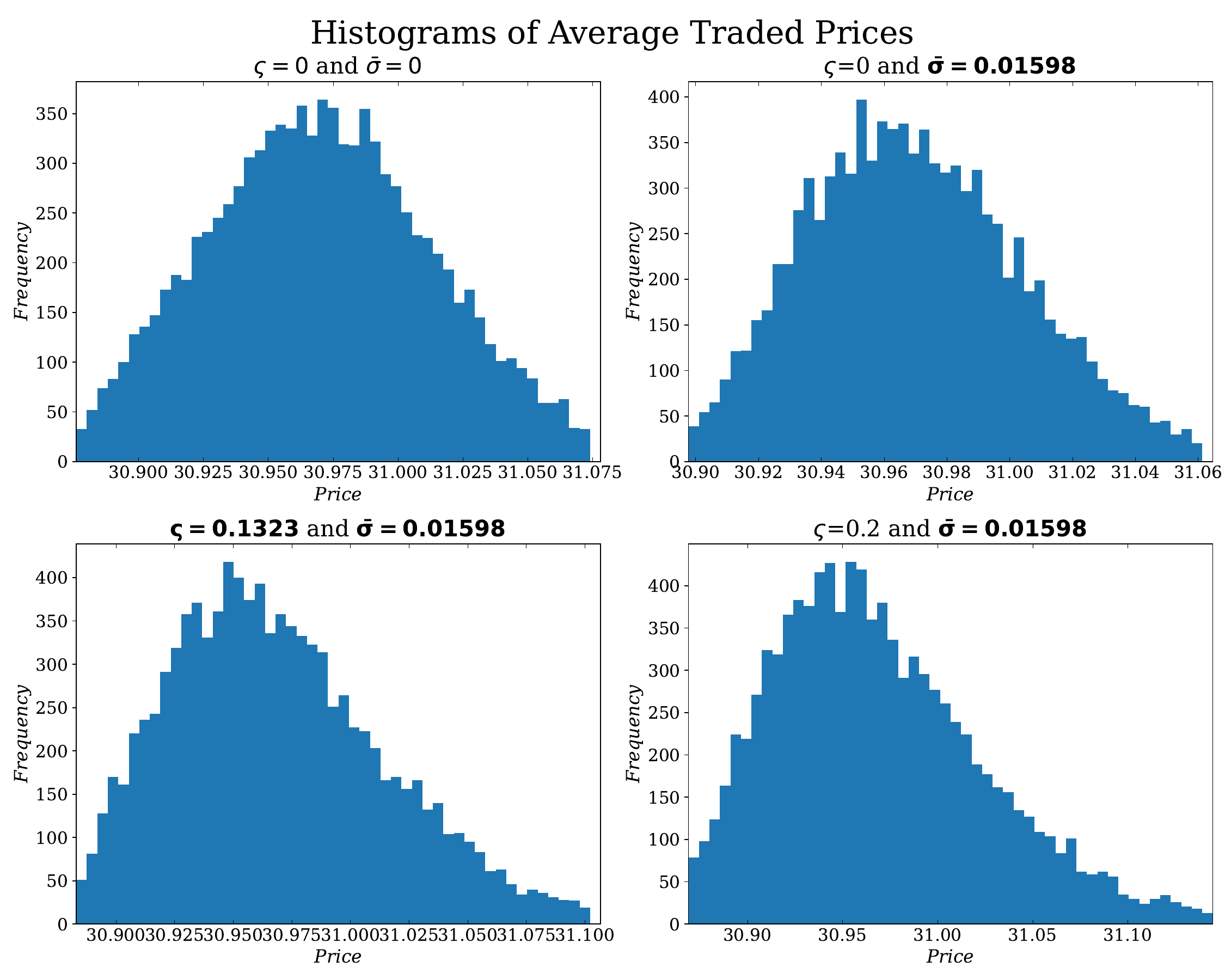}
	\caption{The average liquidation prices over all 10,000 simulations for varying values of either $\bar{\sigma}$ or $\varsigma$.}
	\label{fig:price_histograms_liq}
\end{figure}

\begin{figure}[H]
	\centering
	\includegraphics[width=\linewidth]{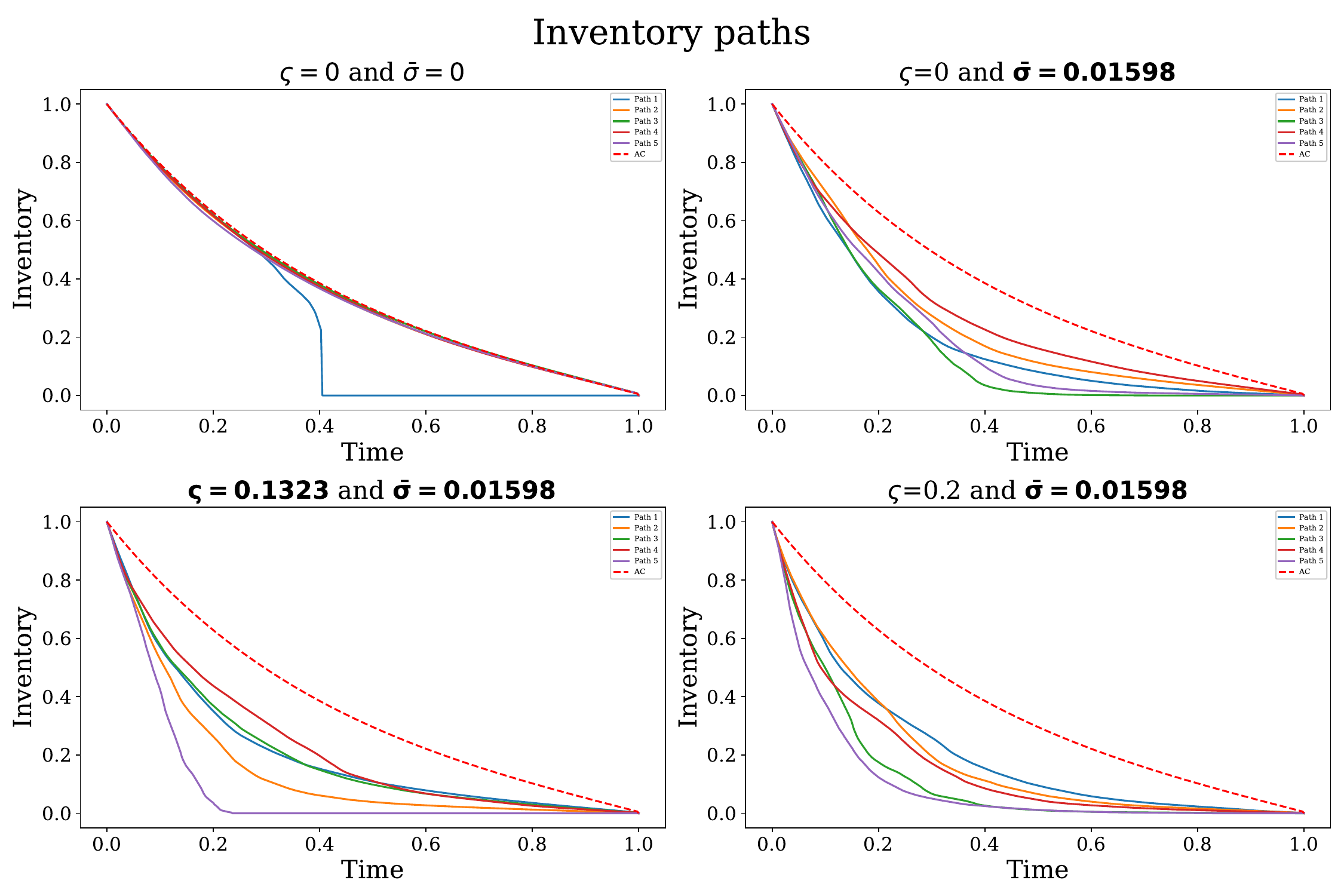}
	\caption{The inventory paths for the five sample price paths over varying values of either $\bar{\sigma}$ or $\varsigma$.}
	\label{fig:inventory_paths_liq}
\end{figure}

\begin{figure}[H]
	\centering
	\includegraphics[width=\linewidth]{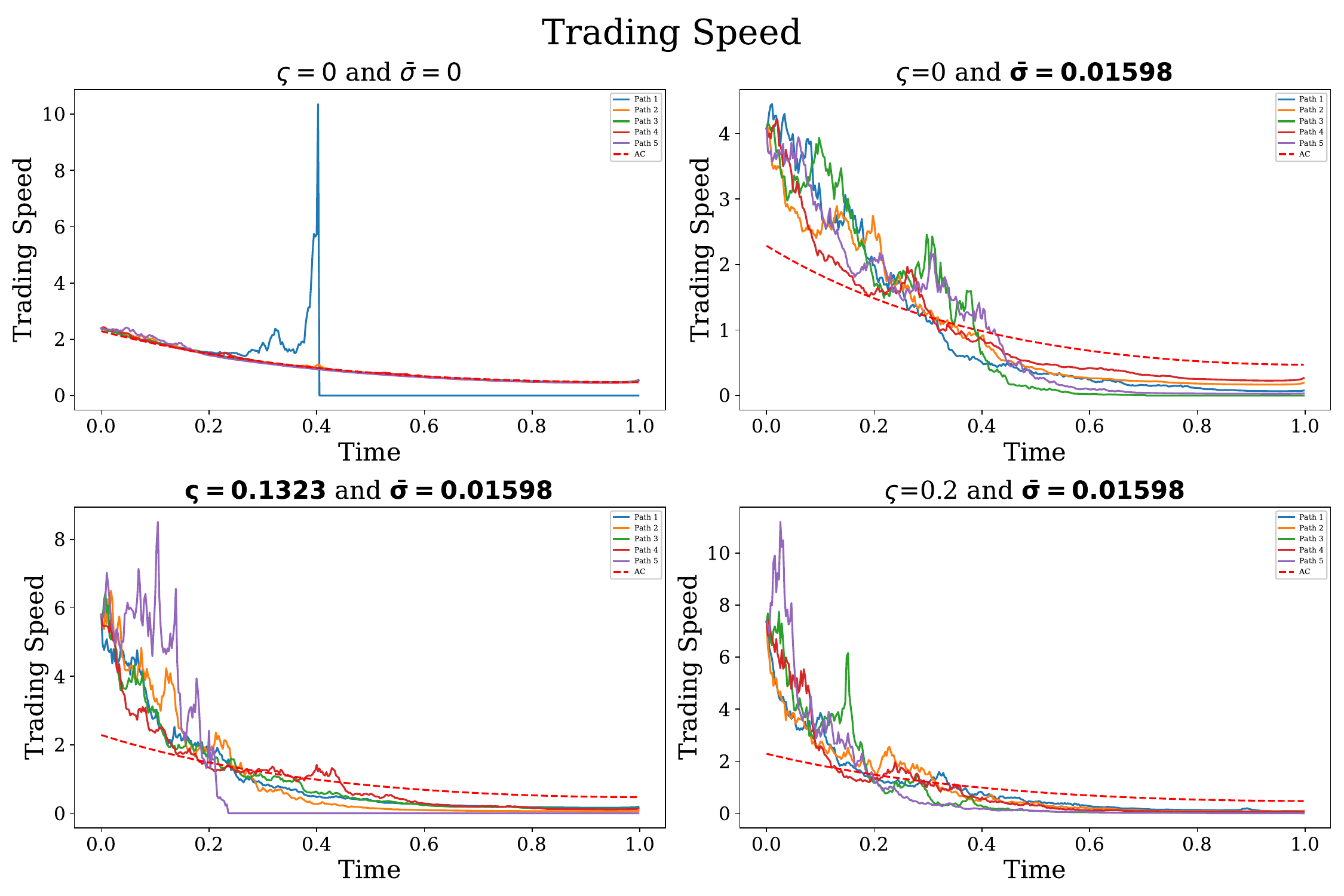}
	\caption{The trading speed paths for the five sample price paths over varying values of either $\bar{\sigma}$ or $\varsigma$.}
	\label{fig:trading_speed_liq}
\end{figure}

Next, in Figures \ref{fig:inventory_paths_liq} and \ref{fig:trading_speed_liq}, we show how the agents inventory and trading speed processes evolve over the five sample price paths we showed above in Figure \ref{fig:price_paths_liq}. These are again the discretized versions of the continuous-time processes given in Equations \eqref{eq:control_h_liq} and \eqref{eq:inv_h_liq}. In these Figures, we again plot the Almgren-Chriss (AC) strategy, where the formula is slightly different to Equation \eqref{eq:trading_speed_AC_acq}, as we are now dealing with the liquidation problem. This formula, as defined in \cite{cartea2015algorithmic}, is,
\begin{align}
\nu_t^{AC} = \sqrt{\kappa\phi}\frac{1+ \xi e ^{2\gamma(T-t)}}{1-\xi e ^{2\gamma(T-t)}},
\end{align}
where $\xi = \frac{\alpha + \sqrt{\kappa\phi}}{\alpha - \sqrt{\kappa\phi}}$ and $\gamma = \sqrt{\frac{\phi}{\kappa}}$, just as for Equation \eqref{eq:trading_speed_AC_acq}. This again holds as long as $\phi >0$, which we defined earlier as the running inventory penalty.

The common theme throughout these subplots is that as we increment either $\bar{\sigma}$ or $\varsigma$, these processes themselves become increasingly more volatile and are heavily determined by the direction and the degree to which the price is moving along each path. We notice in the top left subplot, that the trading strategy (viewed through the lens of the inventory and trading speed processes) does not vary too much along those different price paths. But as we again introduce nonzero increasing values for either of the coefficients $\bar{\sigma}$ or $\varsigma$ from our models, a large portion of the simulations end up liquidating units of the asset a lot faster and many have liquidated the targeted inventory well before the maturity time, $T$. This is not too surprising, as the risk of obtaining good liquidation prices decreases as these coefficients increase, thus the agent is induced to trade a lot faster as a result of this added risk. This is quite important because if a trader were to run a liquidation strategy like this in live markets, and the price were to suddenly jump down, they would want the algorithm to react accordingly. In other words, if a sudden jump down were to enter the market, $\varsigma$ would instantly increase, inducing the trader to increase the speed of liquidating units before the price floor is breached. This increase in the speed of trading would be higher the closer the agent is to the maturity time, $T$. Remember that if prices were to jump up, this would actually be beneficial to the trader, as they would then end up liquidating at higher prices and there should be no increase in urgency to liquidate units of the asset quickly. Thus, we don't focus much on this case as the optimal trading speed is then mostly based on the time left to maturity, which would be quite similar to the AC model for the same number of remaining units of the asset left to liquidate.

\begin{figure}[H]
	\centering
	\includegraphics[width=0.85\linewidth]{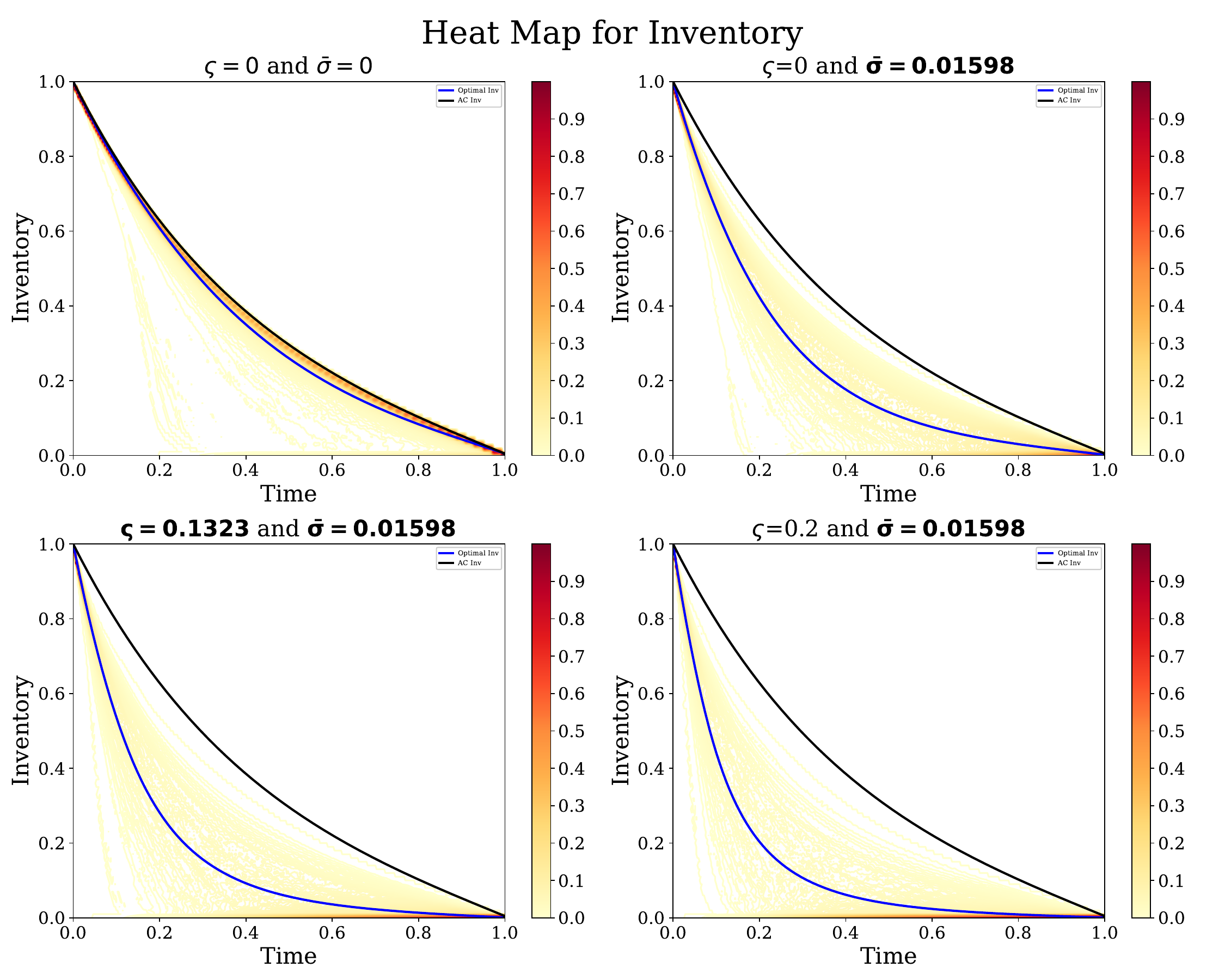}
	\caption{A heat map for the inventory paths over all 10,000 sample price paths over varying values of either $\bar{\sigma}$ or $\varsigma$.}
	\label{fig:inventory_heat_map_liq}
\end{figure}

\begin{figure}[H]
	\centering
	\includegraphics[width=0.85\linewidth]{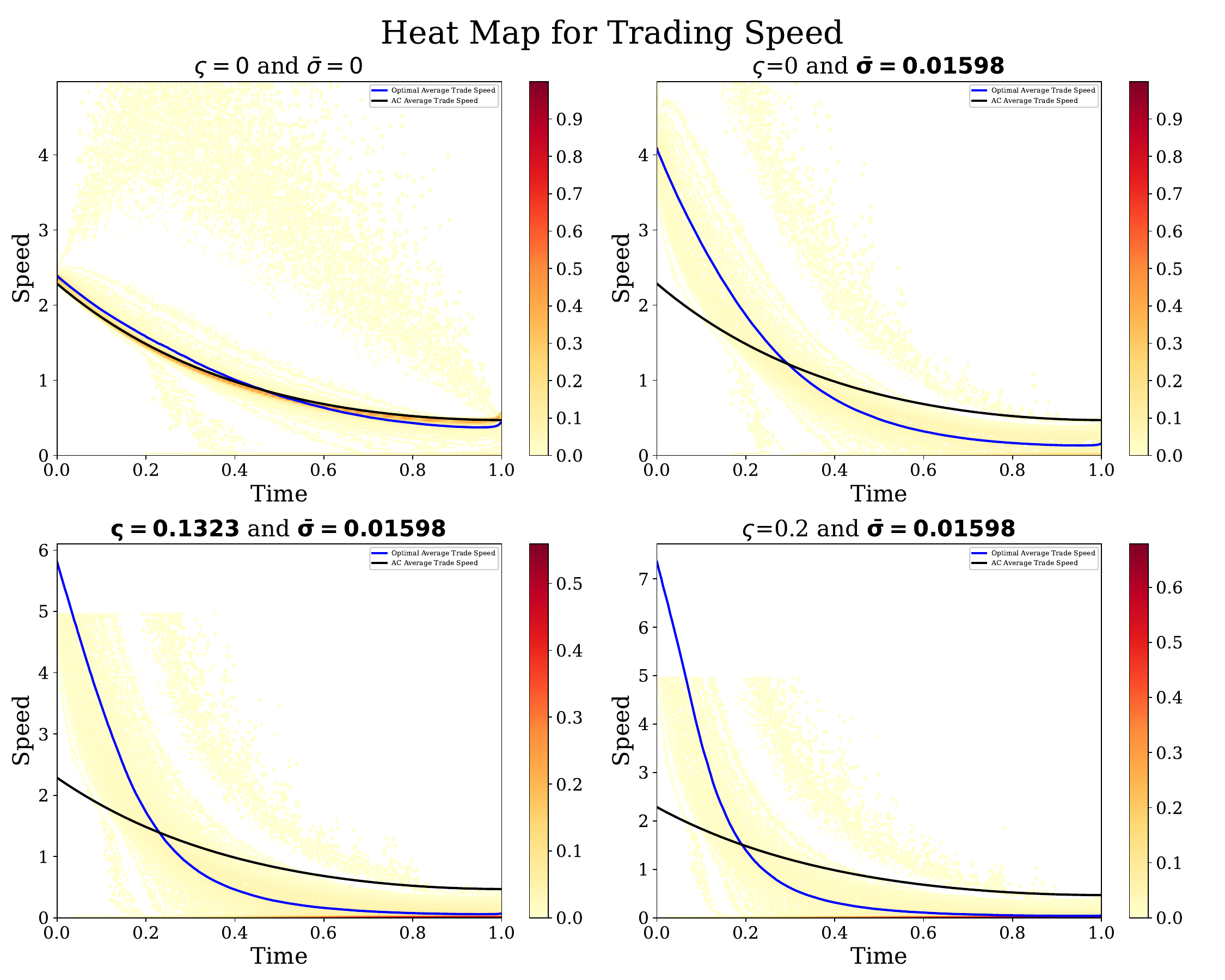}
	\caption{A heat map for the trading speed paths over all 10,000 sample price paths over varying values of either $\bar{\sigma}$ or $\varsigma$.}
	\label{fig:trading_speed_heat_map_liq}
\end{figure}

Lastly, in Figures \ref{fig:inventory_heat_map_liq} and \ref{fig:trading_speed_heat_map_liq}, we show heat maps for the inventory and trading speed processes over all 10,000 simulations. Like in Figure \ref{fig:inventory_heat_map} and \ref{fig:trading_speed_heat_map} for the acquisition strategy simulations, the blue line in both graphs shows the mean of all the inventory and trading speed paths under our optimal strategy, while the black line shows the mean of the inventory and trading speed paths under the AC strategy, which recall is a benchmark strategy. In Figure \ref{fig:inventory_heat_map_liq}, we notice that as we increment either of the parameters $\bar{\sigma}$ and $\varsigma$, the optimal inventory line becomes more convex shaped and thus moves away from the AC line, with this effect increasing the further we are from $t=0$ and $t=T$. In Figure \ref{fig:trading_speed_heat_map_liq}, which is very similar to Figure \ref{fig:trading_speed_heat_map} for the acquisition problem simulations, one will notice that as we increment either of the parameters $\bar{\sigma}$ and $\varsigma$, the trading speed will increase significantly earlier on in the simulation, and thus decreases later on as it liquidated most of its targeted inventory a lot earlier. In both plots, we can see that there are a lot of points at the bottom. These are the simulations where the barrier at $S_{min}$ or $q=0$ got breached before $t=T$. There are more of these instances as we increment either of the parameter values $\bar{\sigma}$ and $\varsigma$. This makes sense intuitively, since when the overall level of volatility is higher, just like in the acquisition problem, the agent is induced to complete the liquidation strategy a lot faster. 

\section{Conclusions and Future Recommendations}
In this paper we introduced a new set of more general price processes that could be implemented in a number of algorithmic and HFT problems. The type of trading problems we focused on apply stochastic optimal control theory, where we chose an example acquisition problem from \cite{cartea2015algorithmic}, as well as designing a similar liquidation problem, for illustrative purposes. These are are common execution trading problems for large institutional investors, who also often measure performance based on how well a large trade was executed.  Our results show that increasing the effect of the jump part (via a diffusion approximation) can significantly alter the optimal trading solution, thus inducing an agent to act differently in the presence of jumps in the asset price. The jump part in our price processes can be either a function of a Semi-Markov or Hawkes process, which have been proven to more accurately mimic LOB dynamics. We provide visuals for how the optimal solution changes based on the effects of jumps, as well as strategy simulations. The strategy simulations portray how the agent would act in different scenarios, based on different random paths of our price processes. Average traded prices, as shown in Figures \ref{fig:price_histograms} and \ref{fig:price_histograms_liq} for the acquisition and liquidation problems, respectively, are a heavily examined measure by institutions for the performance of trading algorithms focused on acquiring or liquidating large amounts of an asset. Thus, under our new models, the main implication is how a trading agent should optimally run specific execution trading algorithms like for when dealing with large positions, based on the more general pricing dynamics given in our Jump-Diffusion model. It is also clear from our analysis that the evolution of the acquisition and liquidation trading strategies, via simulations of the inventory and trading speed processes, can differ significantly based on this additional jump part, which are likely to appear in more volatile market regimes. We would expect a similar effect across a wide range of trading problems using this type of SOC format, with some other examples including trading problems related to market-making, pairs trading, or statistical arbitrage. In-depth descriptions of these types of trading problems in the SOC framework can be found in \cite{cartea2015algorithmic}. 

Regarding future research, we suggest that the simulations in Section 5 be extended to encompass a wider range of market regimes. This could involve changing the parameter values in line with different market-regime rationals and then analyzing the performance of the acquisition and liquidation algorithms. We would also recommend applying similar logic to other types of trading problem setups not limited to the SOC framework used in this paper. For example, reinforcement or deep reinforcement learning (RL) has become very popular lately and is becoming a state-of-the-art method for solving these types of trading problems. \cite{gavsperov2022deep} is one of the first to apply this type of innovative model in a deep RL setting under a Hawkes process pricing model, however, their model is just based on a jump case and no diffusion. Thus, we think this could be an interesting area to combine the two i.e., solve a deep RL trading problem under a jump-diffusion price process, where the jumps are either a function of a Semi-Markov or Hawkes process. We also believe it would be interesting to extend these price models to the Levy case, for cases where the LOB might follow a Markov process (since Semi-Markov and Hawkes processes are not Markov processes). Utilizing Levy processes would enable the use of a more general setting for price processes and would depend on the jump measure. However, we would like to point out that within the SOC framework for the acquisition and liquidation problems we studied in this paper, without our diffusion approximations, the problem would not be solvable under the numerical method we followed similarly to \cite{cartea2015algorithmic}, as you would then be unable to follow the same dimension reduction method. An RL framework may be better suited for trading problems involving advanced price processes, as it can help mitigate these limitations more effectively.

\section*{Acknowledgments}
The authors thank MITACS and NSERC for research funding, and the reviewers for their helpful and insightful comments. 

\section*{Declaration of Interest}
The authors declare no conflict of interest.

\bibliographystyle{apalike}
\bibliography{Paper1}

\end{document}